\documentclass[prx,aps,amssymb,twocolumn]{revtex4-1}
\usepackage{amsmath}
\usepackage{amssymb}
\usepackage{amsthm}
\usepackage{amsfonts}
\usepackage{listings}
\usepackage{enumerate}
\usepackage{latexsym}
\usepackage{color}
\usepackage{bm}
\usepackage{hyperref}
\hypersetup{
 pdfnewwindow=true, colorlinks=true,
 linkcolor=blue, anchorcolor=blue,
 citecolor=blue, filecolor=blue,
 menucolor=blue, urlcolor=blue}

\usepackage{psfrag}

\usepackage{bm}
\usepackage{graphicx}
\usepackage{subfigure}

\usepackage{multirow}

\RequirePackage[normalem]{ulem} 
\RequirePackage{color}\definecolor{RED}{rgb}{1,0,0}\definecolor{BLUE}{rgb}{0,0,1}

\input{epsf}

\begin{document}

\tolerance 10000

\newcommand{\vk}{{\bf k}}

\draft

\title{Ferroelectric metals in 1T/1T$^{\prime}$-phase \\ transition metal dichalcogenide $M$Te$_2$ bilayers ($M =$ Pt, Pd, and Ni)}

\author{Haohao Sheng}
\affiliation{Beijing National Laboratory for Condensed Matter Physics,
and Institute of Physics, Chinese Academy of Sciences, Beijing 100190, China}
\affiliation{University of Chinese Academy of Sciences, Beijing 100049, China}

\author{Zhong Fang}
\affiliation{Beijing National Laboratory for Condensed Matter Physics,
and Institute of Physics, Chinese Academy of Sciences, Beijing 100190, China}
\affiliation{University of Chinese Academy of Sciences, Beijing 100049, China}

\author{Zhijun Wang}
\email{wzj@iphy.ac.cn}
\affiliation{Beijing National Laboratory for Condensed Matter Physics,
and Institute of Physics, Chinese Academy of Sciences, Beijing 100190, China}
\affiliation{University of Chinese Academy of Sciences, Beijing 100049, China}

\begin{abstract}
Ferroelectricity and metallicity cannot coexist due to the screening effect of conducting electrons, and a large number of stable monolayers with 1T/1T$^{\prime}$ phase lack spontaneous polarization due to inversion symmetry. In this work, we have constructed the $\pi$-bilayer structures for transition metal dichalcogenides ($M$Te$_2,M =$ Pt, Pd, and Ni) with van der Waals stacking, where two monolayers are related by $C_{2z}$ rotation, and have demonstrated that these $\pi$ bilayers are typical ferroelectric metals (FEMs). The $\pi$-bilayer structure widely exists in nature, such as 1T$^{\prime}$/T$_d$-TMD, $\alpha$-Bi$_4$Br$_4$. The computed vertical polarization of PtTe$_2$ and MoTe$_2$ $\pi$ bilayers are 0.46 and 0.25 pC/m, respectively. We show that the switching of polarization can be realized through interlayer sliding, which only requires crossing a low energy barrier. The interlayer charge transfer is the source of both vertical polarization and metallicity, and these properties are closely related to the spatially extended Te-$p_z$ orbital. Finally, we reveal that electron doping can significantly adjust the vertical polarization of these FEMs in both magnitude and direction. Our findings introduce a class of FEMs, which have potential applications in functional nanodevices such as ferroelectric tunneling junction and nonvolatile ferroelectric memory.
\end{abstract}

\maketitle
\section{INTRODUCTION}
Ferroelectric (FE) materials, where spontaneous polarization can be reversed or changed by an external electric field~\cite{PhysRev.17.475}, are gradually being developed as two-dimensional (2D) atom-thin layers in order to meet the needs of device miniaturization~\cite{ACSaelm.2019, d1T-MoTe2.NC, In2Se3.NC, SnSe.PRL, AdMa.2021, NbN.PRL}. 
However, many layered transition metal dichalcogenides (TMD) with 2H, 1T, and 1T$^{\prime}$ phases~\cite{TMD.NRM, TMD.CSR} and 2D hexagonal planar structures~\cite{graphene.Science, BN.AFM} have mirror/glide symmetry or inversion symmetry (IS), which precludes spontaneous vertical polarization.
Recently, the concept of sliding ferroelectricity has been proposed~\cite{Nano.2017}, which has successfully expanded the 2D FE family and has opened up a research field known as slidetronics~\cite{MoA2N4.JMCA, npj.8.138, LaBr2.npj, FeCl2.npj, VS2.PRL, BN.Science1, BN.Science2, H-MX2.NatNan, MoS2.NC, MoS2.Nature2022, WTe2.Nature, WTe2.JPCL, WTe2.NP, ReS2.PRL, T1-MoTe2.Nature}. 
This sliding ferroelectricity generates vertical polarization by van der Waals (vdW) stacking, with switching coupled with interlayer sliding.
It has been studied theoretically and experimentally in many materials, such as an h-BN bilayer~\cite{BN.Science1, BN.Science2}, 1H-MoS$_2$ multilayers~\cite{H-MX2.NatNan, MoS2.NC, MoS2.Nature2022}, and 1T$^{\prime}$-WTe$_2$ and ReS$_2$ multilayers~\cite{WTe2.Nature, WTe2.JPCL, WTe2.NP, ReS2.PRL}. The general theory of bilayer stacking ferroelectricity has recently been proposed, which not only includes sliding ferroelectricity but also provides alternative perspectives~\cite{Stacking.arXiv}.

The coexistence of ferroelectricity and metallicity is generally not possible because conducting electrons can effectively screen internal electric fields, preventing the material from exhibiting FE behavior. However, the concept of ferroelectric metal (FEM) has been proposed~\cite{Anderson1965}, and some materials have been proposed as FEM candidates, such as LiOsO$_3$~\cite{LiOsO.NM, LiOsO3.PhysRevB.90.094108}, SrTiO$_3$~\cite{SrTiO.supe.NP, SrTiO.supe.NC}, and Hg$_3$Te$_2X_2$ ($X$ = Cl, Br)~\cite{HgTeX.NC}, but lack direct evidence of polarization switching. 
It has been reported that the polar ($\gamma$; T$_d$ phase) and nonpolar structures ($\bar\gamma$; T$_0$ phase) of metallic MoTe$_2$ can be straightforwardly connected through interlayer shifts~\cite{MoTe2.wzj.PRL, MoTe2-T0.NC}.
In 2D materials, an external electric field can penetrate the material and reverse vertical polarization due to electron confinement within the plane~\cite{CrN.PRB, LiOsO3film.PRL}. 
Despite theoretical predictions of FEM behavior in other 2D materials, such as buckled CrN~\cite{CrN.PRB}, LiOsO$_3$ films~\cite{LiOsO3film.PRL}, and the bimetal phosphates M$_{\rm I}$M$_{\rm II}$P$_2$X$_6$ family~\cite{M1M1P2X6.SB}, only 1T$^{\prime}$-WTe$_2$ multilayers have been experimentally confirmed as FEMs~\cite{WTe2.Nature, WTe2.NP}, and the tunability of polarization by carriers doping in FEMs has been rarely studied yet due to the lack of FEM candidates~\cite{WTe2.doping.NC}.

In this work, we propose that the PtTe$_2$-family $\pi$ bilayers are typical 2D FEMs with spontaneous vertical polarization. Starting from centrosymmetric insulating 1T monolayers, we obtain the metallic $\pi$-bilayer polar structures through vdW stacking. By allowing for interlayer sliding, two FE phases with opposite polarization can switch between each other with low energy cost. The polarization and metallicity of the bilayers can be attributed to charge density redistribution of the Te-$p_z$ orbital induced by interlayer vdW interactions. Finally, we show that the vertical polarization can be significantly adjusted in both magnitude and direction by doping the bilayers with I atoms as electron dopants.

\begin{figure*}[!htb]
\centering
\includegraphics[width=15.75 cm]{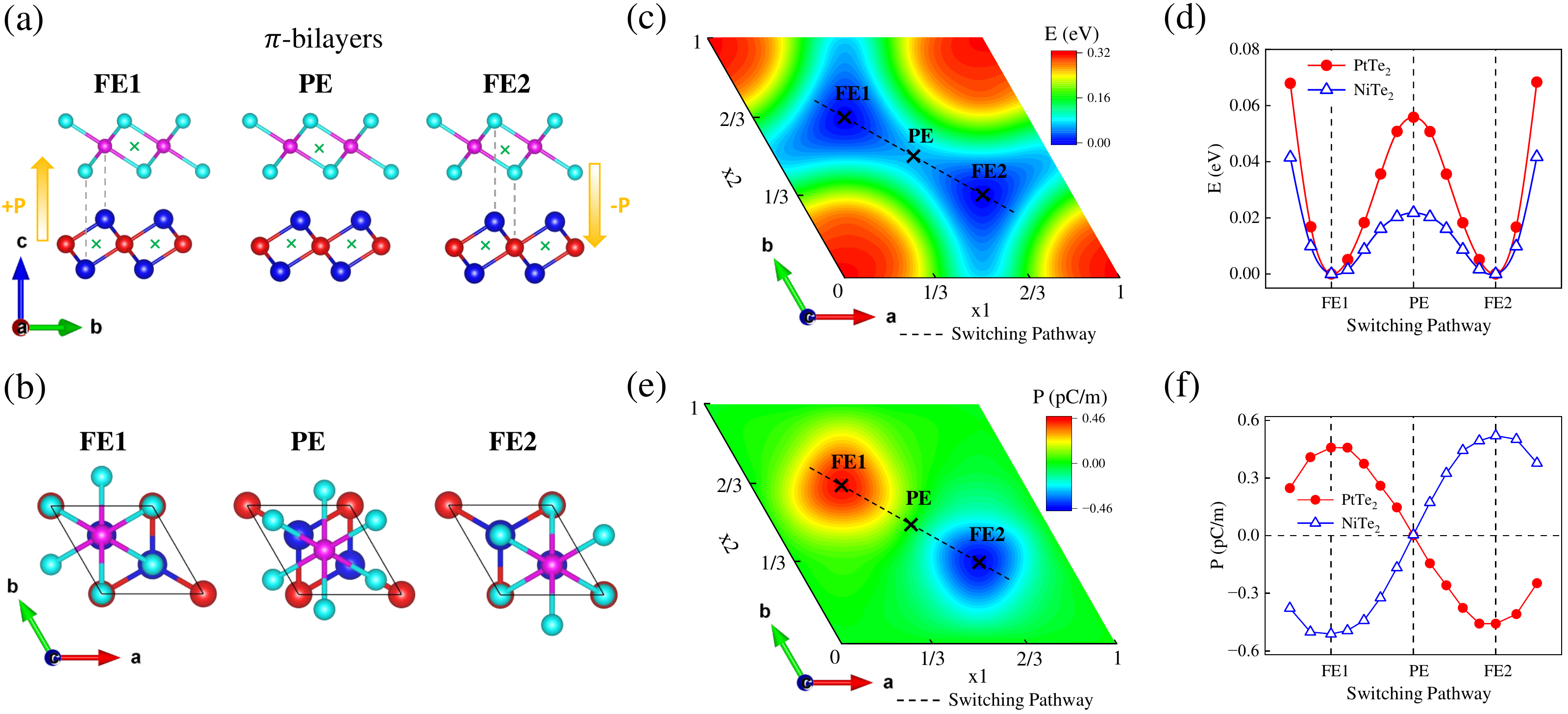}
\caption{(Color online)
(a) Side and (b) top views of two energy-degenerate FE phases (FE1 and FE2) linked by $m_z$ and PE phase for PtTe$_2$ $\pi$ bilayers. The big blue (red) and small cyan (pink) balls represent Te (Pt) atoms in the lower and upper layer, respectively. The green cross represents the inversion center of monolayers. The orange arrows indicate the direction of polarization. Contour plots of (c) total energy and (e) vertical polarization versus the $\textbf{v}_{||}$ for the PtTe$_2$ $\pi$ bilayer. The FE switching pathway is marked with a black dotted line. (d) Energy barriers and (f) polarization transitions on the FE switching pathway for PtTe$_2$ (red circle) and NiTe$_2$ (blue hollow triangle) $\pi$ bilayers. The total energies corresponding to two FE phases are set to zero.
}\label{fig:1}
\end{figure*}

\section{CALCULATIONS AND RESULTS}

\subsection{Crystal structure and vdW stacking}
The 1T-$M$Te$_2$ monolayer possesses a hexagonal structure with space group $P\overline{3}m1$ (SG \#164), where each $M$ atom lies at the center of an octahedral cage formed by Te atoms. IS forbids spontaneous polarization. As reported in Ref~\cite{MoTe2.wzj.PRL}, the two adjacent layers are related by $C_{2z}$ rotation, and the only difference between T$_d$ and T$_0$ phase MoTe$_2$ is the interlayer shift, yielding ferroelectricity in the semimetal T$_d$-MoTe$_2$.
We construct a bilayer system, where the top and bottom 1T monolayers ($xy$ plane) are related by $C_{2z}$ rotation. The calculation details are given in Appendix~\ref{SM1}.
Hereafter, starting from the centrosymmetric monolayer without $C_{2z}/M_z$ symmetry, the $C_{2z}$-related bilayer system is called the $\pi$ bilayer, which breaks IS. 
The $\pi$-bilayer structure can be widely found in nature materials, such as 1T$^{\prime}$/T$_d$/T$_0$-TMD, $\alpha$-Bi$_4X_4$ ($X$=Br, I).
Note that the monolayer of the T$_d$ and T$_0$ phase is identical.
Exfoliated from the T$_d$-MoTe$_2$ bulk, we calculate the FE polarization of the metallic MoTe$_2$ $\pi$ bilayer being 0.25 pC/m, where FE switching under the electric field can be easily achieved (Appendix~\ref{SM2}).
In addition, the crystal structures, $\pi$ bilayers, and vertical polarization of topological compounds Bi$_4X_4$ are presented in detail in Appendix~\ref{SM2}. 

In the series of $\pi$-bilayer structures, a glide symmetry ($g_z$: mirror symmetry $m_z$ with a fractional in-plane translation) is preserved in the systems with $x_{1,2} = 0, 1/2$, while others with $x_{1,2}\neq 0 ~or~1/2$ do not have such a $g_z$. 
The interlayer Te atoms distance $d_z$ of the $\pi$ bilayer is obtained by full relaxation. Its structure is parametrized by the $\textbf{v}_{||}\equiv M_{top}-M_{bottom}=(x_1,x_2)$ (with respect to the lattice constants; $\textbf{v}_{||}=x_1 \textbf{a} +x_2 \textbf{b}$).
The two FE phases with SG $P3m1$ (\#156) of PtTe$_2$ are obtained in our calculations [FE1: $\textbf{v}_{||}=(1/3,2/3)$; FE2: $\textbf{v}_{||}=(2/3,1/3)$], whose structures are shown in Figs.~\ref{fig:1}(a) and \ref{fig:1}(b). 
Because the two FE structures are related by $m_z$, they have the opposite vertical polarizations.
The vertical polarizations of $M$Te$_2$ FE $\pi$ bilayers are presented in Table~\ref{table:1}.
For the PtTe$_2$ $\pi$ bilayer, the vertical polarization is 0.46 pC/m, comparable to that of the previous FEM WTe$_2$ ($\sim$0.40 pC/m)~\cite{WTe2.Nature, WTe2.JPCL}.
Due to the existence of $C_{3z}$ symmetry, the net in-plane polarization is zero.
The existence of vertical polarization is also confirmed by the discontinuity of vacuum levels on the upper and lower sides of the slab, details of which are shown in Appendix~\ref{SM3}.
These results are consistent with the general theory for bilayer stacking ferroelectricity~\cite{Stacking.arXiv}.

\begin{figure*}[!htb]
\centering
\includegraphics[width=17.5 cm]{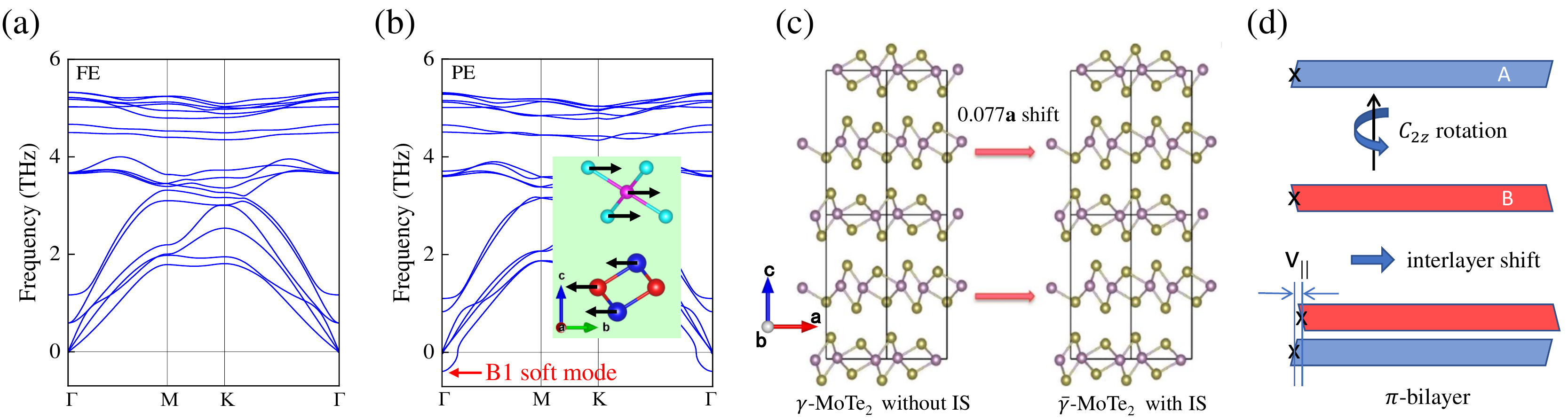}
\caption{(Color online)
Phonon spectra of (a) FE and (b) PE phases for PtTe$_2$ $\pi$ bilayers. Inset presents the vibration pattern of optical soft mode at $\Gamma$ point with B1 irreducible representation for PE phase. PE phase is unstable and can spontaneously transition to two FE phases.
(c) The T$_d$ phase ($\gamma$) and T$_0$ phase ($\bar\gamma$) of MoTe$_2$ are straightforwardly connected through interlayer shifts. The two centrosymmetric 1T$^{\prime}$ monolayers are related by $C_{2z}$ rotation, and the interlayer shifts break IS in the bulk crystals (adapted from Ref.~\cite{MoTe2.wzj.PRL}).
(d) Schematic diagram of the $\pi$-bilayer structure. Two centrosymmetric monolayers are related by $C_{2z}$ rotation and interlayer shift. The black cross represents the inversion center of monolayers. The 1T$^{\prime}$/T$_d$-MoTe$_2$ and $\alpha$-Bi$_4$Br$_4$ contain the $\pi$-bilayer structure in nature. The vertical polarizations of MoTe$_2$ and Bi$_4$Br$_4$ $\pi$ bilayers are computed to be 0.25 pC/m and 0.05 pC/m, respectively.
} \label{fig:2}
\end{figure*}

\begin{table}[!htb]
\begin{ruledtabular}
\caption{Vertical polarizations (P: pC/m) of the FE1 phase, interlayer Te atoms distance ($d_0$: \AA), and energy barriers (E$_{barrier}$: meV per unit cell) of the FE switching pathway for $M$Te$_2$ FE $\pi$-bilayers.}
   \begin{tabular}{c c c c }%
          FE1      & P          & $d_0$      & E$_{barrier}$  \\
            \hline
            PtTe$_2$     &  ~0.46        & 2.67       &  56           \\
            PdTe$_2$     &  ~0.06        & 2.54       &  32            \\
            NiTe$_2$     & -0.52         & 2.66       &  22             \\
  \end{tabular}
\label{table:1}
\end{ruledtabular}
\end{table}

\subsection{Ferroelectric switching through interlayer sliding}
From the total energy and vertical polarization as a function of the $\textbf{v}_{||}$ in Figs.~\ref{fig:1}(c) and \ref{fig:1}(e), we obtained an FE switching pathway realized by interlayer sliding as depicted by the dashed line. The energy barrier and polarization transition on the pathway are plotted in Figs.~\ref{fig:1}(d) and \ref{fig:1}(f), respectively. As global ground states, two FE phases are dynamically stable, which is confirmed by phonon spectra in Fig.~\ref{fig:2}(a) for the PtTe$_2$ FE $\pi$ bilayer.
The phonon spectra of the PtTe$_2$ paraelectric (PE) $\pi$ bilayer is presented in Fig.~\ref{fig:2}(b), with the vibration pattern of the B1 optical soft mode at $\Gamma$ point depicted in the inset.
The PE phase [SG Aem2 (\#39)] with $\textbf{v}_{||}=(1/2,1/2)$ at the saddle point of the energy surface is unstable and will spontaneously transition to two FE phases, which indicates that the PE phase is the transition state of the FE switching pathway.

In addition, the typical energy double well structure is clearly visible in Fig.~\ref{fig:1}(d), with a low energy barrier of 56 meV per unit cell for the PtTe$_2$ $\pi$-bilayer, which is listed in Table~\ref{table:1}. Such a low energy barrier is attributed to the fact that the interlayer sliding to achieve FE switching only needs to overcome weak interlayer vdW interactions, which does not involve the deformations of tightly bonded atoms as in displacive ferroelectrics~\cite{In2Se3.NC, BaTiO.Nature}. Sliding driven the oscillation of interlayer potential can generate alternating current, which provides a platform for the development of a nanogenerator~\cite{Nano.2017, WTe2.JPCL}.
For PdTe$_2$ and NiTe$_2$ $\pi$ bilayers, contour plots of total energy and vertical polarization versus the $\textbf{v}_{||}$  can be found in Appendix~\ref{SM3}, which are similar to those of the PtTe$_2$ $\pi$ bilayer.

\begin{figure*}[!htb]
\centering
\includegraphics[width=15 cm]{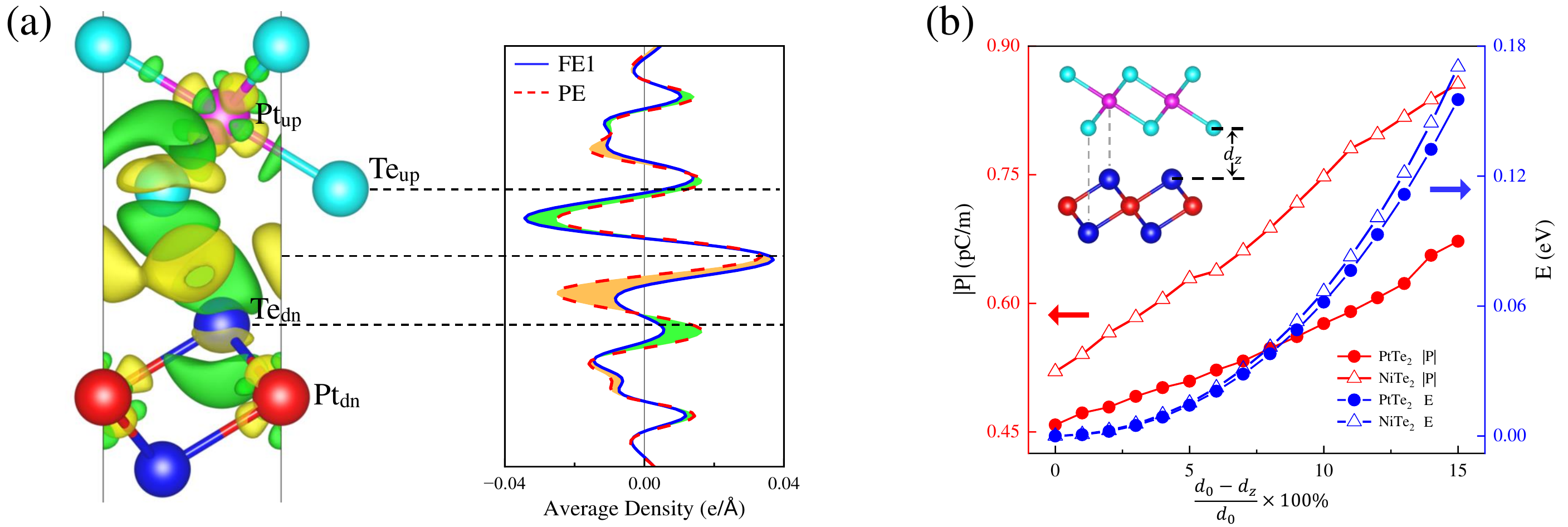}
\includegraphics[width=15 cm]{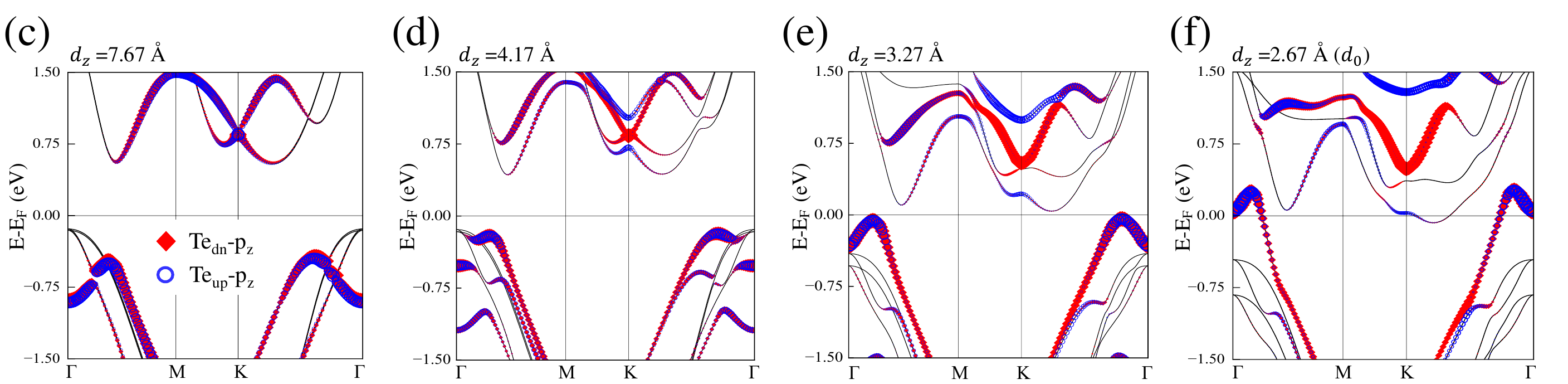}
\caption{(Color online)
(a) Charge density difference of FE1 phase and plane-averaged charge density difference along the $z$ direction of FE1 (solid blue line) and PE (red dotted line) phases for PtTe$_2$ $\pi$ bilayers, in terms of constituent monolayers. The value of the isosurface is set to 0.003 $e/$\AA$^3$. Yellow and green regions represent electron accumulation and depletion, respectively. The black dotted lines mark the $z$ positions of the center of the PtTe$_2$ $\pi$ bilayer and the Te$_{up}$ and Te$_{dn}$ atoms. (b) Interlayer Te atoms distance $d_z$ dependence of vertical polarization and total energy for PtTe$_2$ (circle) and NiTe$_2$ (triangle) FE $\pi$ bilayers. The total energies corresponding to $d_0$ are set to zero. The definition of $d_z$ is shown in the inset and $d_0$ represents the actual $d_z$ obtained by fully relaxation. [(c)-(e)] Orbital-resolved band structures for PtTe$_2$ FE $\pi$ bilayers with (c) $d_z$=7.67 \AA, (d) $d_z$=4.17 \AA, (e) $d_z$=3.27 \AA, and (f) $d_z$=2.67 \AA ($d_0$). The sizes of the red diamonds and blue hollow circles represent the weights of Te$_{dn}$-$p_z$ and Te$_{up}$-$p_z$ orbitals, respectively.
}
\label{fig:3}
\end{figure*}

\subsection{Origin of vertical polarization}
Interlayer sliding can be equivalent to rigid ion translation without vertical displacement, so vertical polarization is purely electronic in origin. We attribute the origin of polarization to the asymmetric charge density redistribution induced by interlayer vdW interactions, instead of the subtle ion displacement along the polarization direction from the inversion center as in displacive ferroelectrics represented by perovskite oxides~\cite{BaTiO.Nature}. In the FE1 phase of Fig.~\ref{fig:3}(a), the Te atom (Te$_{dn}$) at the interface in the lower layer is directly below the Pt atom (Pt$_{up}$) in the upper layer, while the Te atom (Te$_{up}$) at the interface in the upper layer sits above the center of the triangle spanned by the Pt atom (Pt$_{dn}$) in the lower layer. This asymmetric vdW stacking causes strong distortion of the spatially extended Te$_{dn}$-$p_z$ orbital, making the interlayer vdW charge transfer near the Te$_{dn}$ atom significantly deviate from the symmetric distribution of the PE phase, as shown by the plane-averaged charge density difference along the $z$ direction in Fig.~\ref{fig:3}(a) for the PtTe$_2$ $\pi$ bilayer. 
When the top layer slides along the switching pathway, the interlayer vdW charge transfer gradually reverses, resulting in the reversal of vertical polarization. Additionally, FE materials usually have a piezoelectric effect. By compressing $d_z$ in Fig.~\ref{fig:3}(b), vertical polarization can be further enhanced due to more intense vdW charge transfer.

\subsection{Metallicity}
It is worth mentioning that the vdW charge transfer also introduces additional metallicity into our FE $\pi$ bilayers. Although the monolayer is insulating, the bilayer becomes metallic, which is further checked by the HSE06 functional (Appendix~\ref{SM4}). After the stacking, Te$_{up}$ and Te$_{dn}$ atoms lose electrons seriously due to the space expansion of the $p_z$ orbital, forming a large charge depletion zone like the $p_z$ orbital, as shown by the charge density difference in Fig.~\ref{fig:3}(a) for the PtTe$_2$ $\pi$ bilayer. When the two monolayers are far apart in Fig.~\ref{fig:3}(c), the band structure can be regarded as a direct doubling of the band of monolayer (Appendix~\ref{SM5}). With the increase of the interlayer vdW charge transfer due to the reduced $d_z$ in Figs.~\ref{fig:3}(d)-\ref{fig:3}(f), the valence band dominated by Te$_{up}$-$p_z$ and Te$_{dn}$-$p_z$ orbitals gradually rise above $E_F$ near the $\Gamma$ point, moving the PtTe$_2$ FE $\pi$ bilayer from an insulating phase to a metallic phase (see more in Appendix~\ref{SM5}).

In particular, the origin of metallicity of $M$Te$_2$ $\pi$ bilayers is different from that of the previous FEM WTe$_2$ multilayers, where the centrosymmetric monolayer is already metallic~\cite{WTe2.Nature, WTe2.JPCL}. In addition, by stacking 2D FE semiconductors to manipulate the depolarization field, it is possible to drive the inversion of the valence band maximum and the conduction band minimum, which is successful in introducing metallicity into the In$_2X_3$ ($X$ = S, Se, and Te) FE system~\cite{In2Se3.MH}. Electron doping is also an effective strategy. Very recently, by applying external gate bias to introduce free carriers, the coexistence of vertical polarization and in-plane conductance is realized in MoS$_2$ and WSe$_2$ insulated FE bilayers~\cite{MoS2.Nature2022}.

\begin{figure}[!htb]
\centering
\includegraphics[width=6 cm]{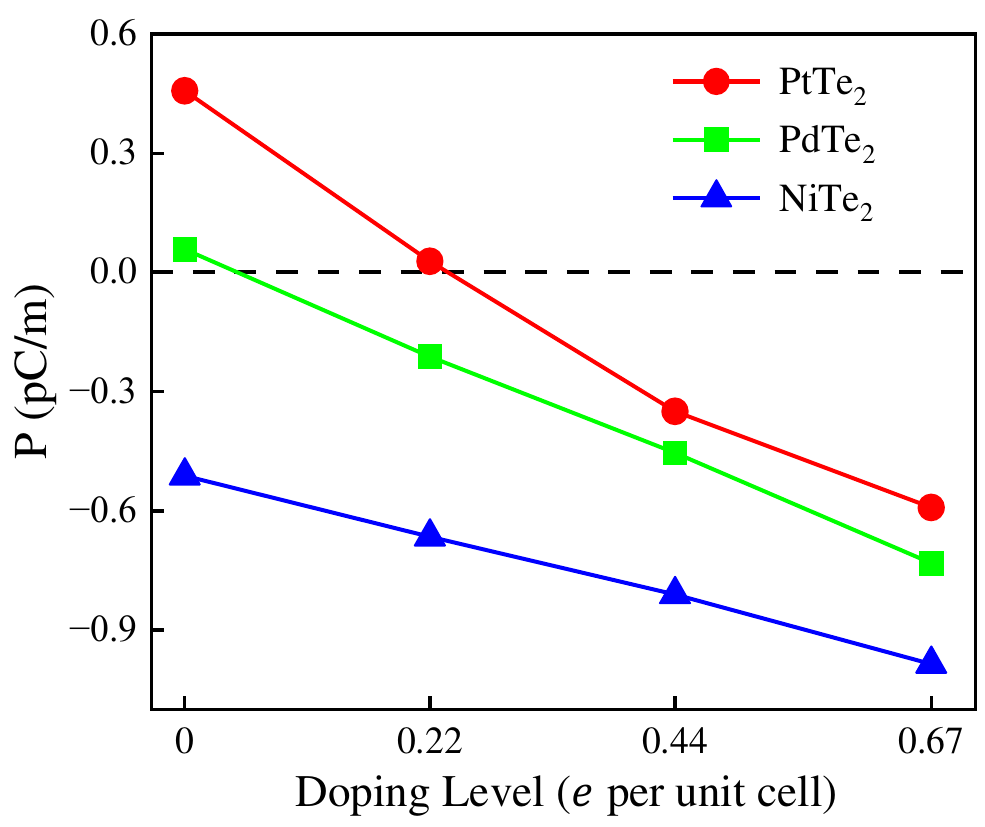}
\caption{(Color online)
Vertical polarization as a function of doping level ($e$ per unit cell) for PtTe$_2$ (red), PdTe$_2$ (green), and NiTe$_2$ (blue) $\pi$ bilayers with FE1 phase.
}
\label{fig:4}
\end{figure}

\subsection{Polarization versus doping}
Next we will discuss the tunability of polarization in PtTe$_2$-family $\pi$ bilayers through electron doping. Due to the metallic nature of FEMs, their carrier densities can be easily modified, and this can be simulated by replacing Te atoms with I atoms as electron doping. In Fig.~\ref{fig:4}, we simulate carrier densities regulation of polarization. The vertical polarization of the PtTe$_2$ $\pi$ bilayer is shown to decrease to zero and then increase in the opposite direction with the increase of additional electrons. The magnitude of polarization reaches a maximum of 0.59 pC/m when doping 0.67 $e$ per unit cell, which is slightly larger than the original polarization of the PtTe$_2$ $\pi$ bilayer. The introduction of effective electron doping in $M$Te$_2$ $\pi$ bilayers causes the negative charge centers to move upward relative to their positive charge centers. 
Recently, the full tunability of polarization by changing carrier densities in the FEM WTe$_2$ bilayer is proved by experiments~\cite{WTe2.doping.NC}.
Our results suggest that this ability to significantly adjust vertical polarization through charge carrier doping is a great advantage of $M$Te$_2$ FEMs and will be of great experimental interest.

\section{DISCUSSION}
FEMs (or FE-like metals) are rare in nature but have demonstrated various fascinating properties, such as unconventional superconductivity~\cite{SrTiO.supe.NP, SC.Nature2004}, unique optical responses~\cite{OP.PRB2010, OP.PRB2011}, and magnetoelectric effects~\cite{ME.PRL1995, ME.PRB2018}. The vdW layered materials provide an ideal platform for the study of sliding and metallic ferroelectricity due to their unique interlayer interactions. In this work, we demonstrate that the series of PtTe$_2$-family $\pi$-bilayers are typical FEMs with significant adjustment of vertical polarization by changing carrier densities. First, we design the metallic $\pi$-bilayer structure with vertical polarization by starting from centrosymmetric insulating 1T monolayers. Second, we show that the switching of vertical polarization is coupled with interlayer sliding, where a low energy barrier ensures high speed and energy-saving data storage and processing. Third, we explain that the polarization and metallicity are caused by charge density redistribution of the Te-$p_z$ orbital, induced by interlayer vdW interactions. Finally, we find that vertical polarization can be significantly adjusted in both magnitude and direction by doping I atoms, which is a advantage of $M$Te$_2$ FEMs. Electron doping causes the negative charge center to move upward relative to the positive charge center. These results predict a class of typical 2D FEMs, which have potential applications in functional nanodevices such as ferroelectric tunneling junction and nonvolatile ferroelectric memory.
In addition, the $\pi$ bilayer of MoTe$_2$ simultaneously exhibits ferroelectricity and superconductivity, which can be tuned by electron or hole doping. 
As the Bi$_4X_4$ monolayer is a quantum spin Hall insulator, the Bi$_4X_4$ $\pi$ bilayer would show both ferroelectricity and topology.
In conclusion, the $\pi$ bilayer structure can be obtained in nature or in laboratory, which enables us to manipulate the interplay between the ferroelectricity and other properties, such as superconductivity and topology.

\section{ACKNOWLEDGMENTS}
We thank Professor Hongjun Xiang for helpful discussions.
This work was supported by the National Key R\&D Program of China (Grant No. 2022YFA1403800), the National Natural Science Foundation of China (Grants No. 11974395 and No. 12188101), the Strategic Priority Research Program of Chinese Academy of Sciences (Grant No. XDB33000000), and the Center for Materials Genome.


\setcounter{section}{0}
\setcounter{subsection}{0}
\setcounter{equation}{0}
\setcounter{table}{0}

\renewcommand{\thesubsection}{\Alph{subsection}}
\renewcommand{\theequation}{A\arabic{equation}}
\renewcommand{\thetable}{A\arabic{table}}

\appendix

\begin{figure}[!htb]
\centering
\includegraphics[width=8.5 cm]{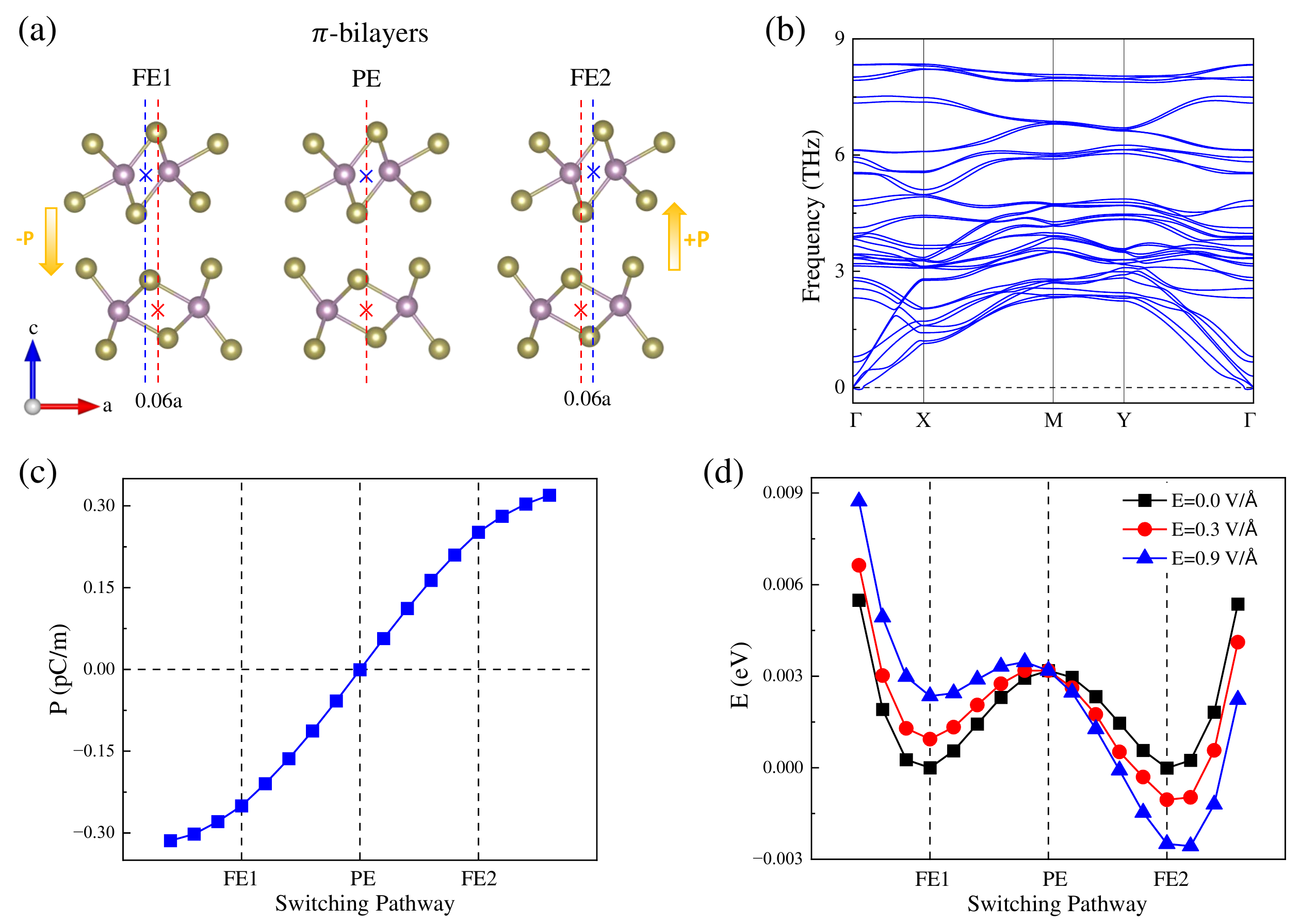}
\caption{(Color online)
(a) Side view of two energy-degenerate FE phases (FE1 and FE2) linked by $m_z$ and PE phase for MoTe$_2$ $\pi$ bilayers. The blue and red crosses represent the inversion center of the monolayer in the upper and lower layer, respectively. The orange arrows indicate the direction of polarization. (b) Phonon spectra of FE phase for the MoTe$_2$ $\pi$ bilayer. 
(c) Polarization transitions and (d) energy barriers (under different vertical electric fields) on the FE switching pathway
for the MoTe$_2$ $\pi$ bilayer. The total energies corresponding to two FE phases without external electric field are set to zero, and the energy curves with external electric field are shifted so that the total energies of PE phases without and with external electric field coincide.
} \label{fig:S1}
\end{figure}

\begin{figure*}[!htb]
\centering
\includegraphics[width=14.5 cm]{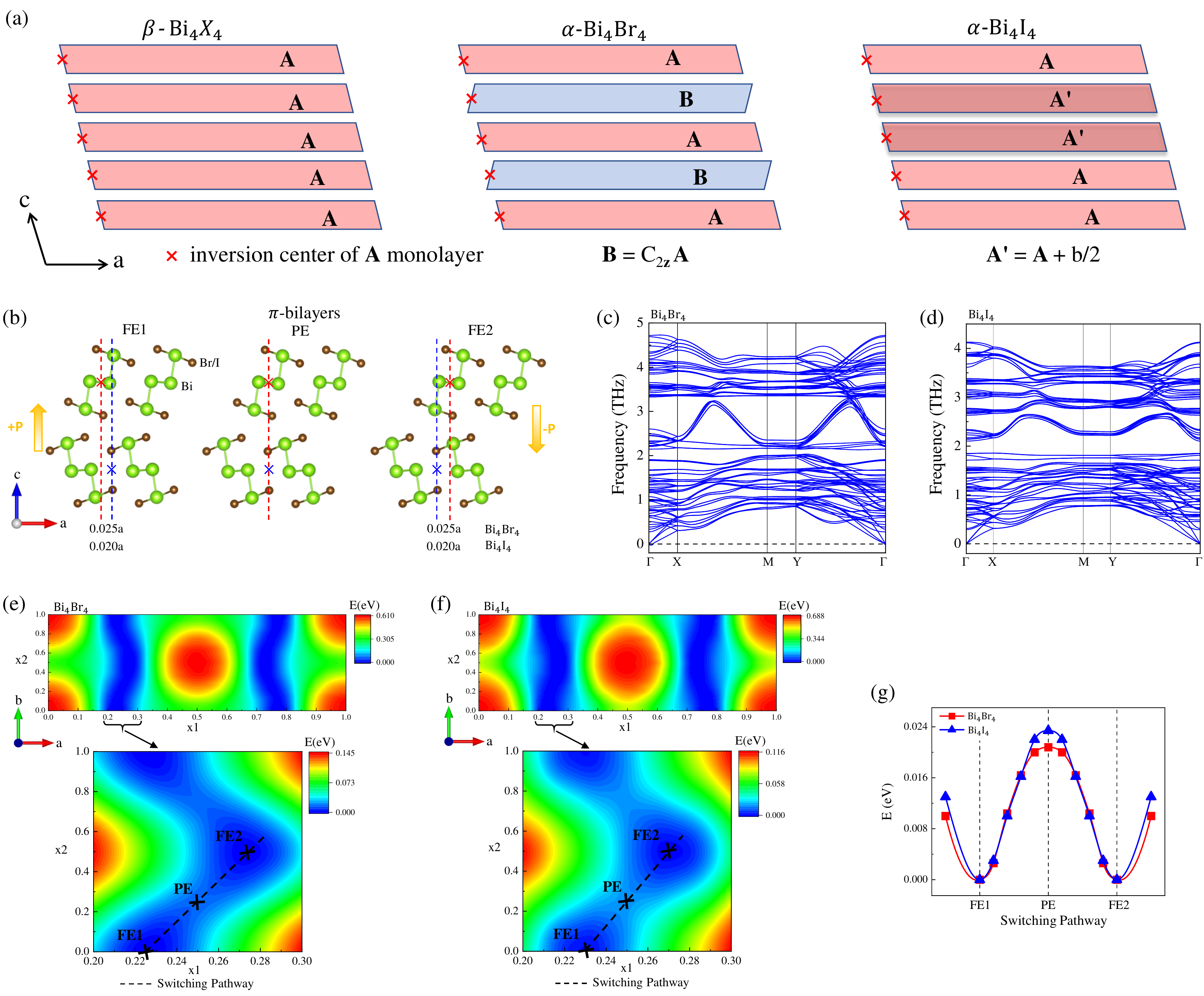}
\caption{(Color online)
(a) Schematic diagram of $\beta$ and $\alpha$-Bi$_4X_4$ ($X$=Br, I) structures.
(b) Side view of two energy-degenerate FE phases (FE1 and FE2) linked by $m_z$ and a PE phase for Bi$_4X_4$ $\pi$ bilayers. 
The orange arrows indicate the direction of polarization.  Phonon spectra of the FE phase for (c) Bi$_4$Br$_4$ and (d) Bi$_4$I$_4$ $\pi$ bilayers. Contour plots of total energy versus the $\textbf{v}_{||}$ for (e) Bi$_4$Br$_4$ and (f) Bi$_4$I$_4$ $\pi$ bilayers. The FE switching pathway is marked with a black dotted line. (g) Energy barriers on the FE switching pathway for Bi$_4$Br$_4$ (red circle) and Bi$_4$I$_4$ (blue triangle) $\pi$ bilayers. The total energies corresponding to two FE phases are set to zero.
} \label{fig:S2}
\end{figure*}

\section{CALCULATION AND METHODOLOGY}
\label{SM1}
We carried out the first-principles calculations based on the density functional theory (DFT) with projector augmented wave (PAW) method~\cite{paw1, paw2}, as implemented in the Vienna \emph{ab initio} simulation package (VASP)~\cite{KRESSE199615, vasp}. The generalized gradient approximation (GGA) in the form of Perdew-Burke-Ernzerhof (PBE) function~\cite{pbe} was employed for the exchange-correlation potential. The kinetic energy cutoff for plane wave expansion was set to 500 eV, and a 18$\times$18$\times$1 Monkhorst-Pack $\textbf{k}$-mesh was adopted for the Brillouin zone sampling in the self-consistent process. 
The first-order Methfessel-Paxton scheme with the width of the smearing 0.1 eV is used as the k-space integration/smearing method.
The thickness of the vacuum layer along the $z$ axis was set to \textgreater ~20 \AA. Both lattice parameters and atomic positions were fully relaxed by minimizing the interionic forces below 10$^{-2}$ eV/\AA. The DFT-D3 method of Grimme with BJ damping was applied to consider vdW interaction~\cite{DFT-D3.JCP,DFT-D3.JCC}. The dipole correction was applied in all calculations~\cite{dipole}.  
The Heyd–Scuseria–Ernzerhof (HSE06) hybrid functional~\cite{HSE.JCP} was employed to check band structure.
An FE switching pathway was obtained with the climbing-image nudged elastic band method~\cite{CINEB}. Phonon spectra were gained with the finite-difference method using a 3$\times$3$\times$1 supercell, as implemented in the Phonopy package~\cite{phonopy}.

Due to the disappearance of polarization uncertainty caused by the absence of periodicity in the out-of-plane direction, the vertical polarization of 2D systems is well defined by the classical dipole method~\cite{In2Se3.NC, WTe2.JPCL, M1M1P2X6.SB, In2Se3.MH}. The ion vertical polarization can be calculated by the point charge model, as shown in the following formula:
\begin{equation}
  P_{ion}=\frac{e}{S}\sum_{i}z_iQ_i=\frac{e}{S}\overline{z}_{ion}n
\end{equation}
where $S$ is the in-plane area of the unit cell, $e$ is the elementary charge, $z_i$ is the spatial position of the $i^{th}$ ion along the $z$ direction, $Q_i$ is the ionic charge of the $i^{th}$ ion, $n$ is the total number of electrons, and $\overline{z}_{ion}$ represents the ion center (positive charge center) of the system in the $z$ direction. The sum is over all ions in the unit cell.  

Because of the continuous distribution of electron cloud, the electron vertical polarization can be calculated using the following equation:
\begin{equation}
  P_e=-\frac{e}{S}\iiint z\rho(\textbf{r})d\textbf{r}=-\frac{e}{S}\overline{z}_{e}n \label{Pe}
\end{equation}
where $\rho(\textbf{r})$ represents the electronic charge density, $\overline{z}_{e}$ represents the electron center (negative charge center) of the system in the $z$ direction, and the integral is over the whole unit cell.

The total vertical polarization is the sum of the ion polarization and the electron polarization,
\begin{equation}
  P=P_{ion}+P_e=\frac{e}{S}\sum_{i}z_iQ_i-\frac{e}{S}\iiint z\rho(\textbf{r})d\textbf{r} \label{Ptotal}
\end{equation}
where the direction is from the negative charge center to the positive charge center.

\section{FERROELECTRIC IN MoTe$_2$ AND Bi$_4X_4$ $\pi$ BILAYERS}
\label{SM2}
The 1T$^{\prime}$-MoTe$_2$ monolayer with SG $P2_1/m$ (\#11) hosts IS and lacks spontaneous polarization. We construct the $\pi$-bilayer structure, which can be directly exfoliated from the 1T$^{\prime}$/T$_d$/T$_0$-phases MoTe$_2$ bulk~\cite{MoTe2.wzj.PRL, MoTe2-T0.NC}, to introduce ferroelectricity.
As shown in Fig.~\ref{fig:S1}(a), the two FE phases with SG $Pm$ (\#6) and the PE phase with SG $Pmc2_1$ (\#26) are obtained in our calculations [FE1: $\textbf{v}_{||}=(0.44,0)$; FE2: $\textbf{v}_{||}=(0.56,0)$; PE: $\textbf{v}_{||}=(0.50,0)$]. 
Because the two FE structures are related by $m_z$ symmetry, they have the opposite vertical polarizations (0.25 pC/m). A glide symmetry ($g_z$: $m_z$ with a fractional in-plane translation) and $C_{2x}$ symmetry in the PE phase forbid vertical polarization. 
As confirmed by phonon spectra in Fig.~\ref{fig:S1}(b), the two FE phases are dynamically stable. The FE switching is realized by interlayer sliding. The polarization transition and energy barrier (under different vertical electric fields) on the FE switching pathway are plotted in Fig.~\ref{fig:S1}(c) and \ref{fig:S1}(d), respectively. The typical energy double well is clearly visible, with a low energy barrier of 3.18 meV per unit cell. In addition, we investigate the influence of the external electric field perpendicular to the slab on the energy pathway of FE switching, where the vertical electric field is applied by introducing a dipole layer in the middle of the vacuum region~\cite{dipole}. With increasing the electric field in Fig.~\ref{fig:S1}(d), the energy barrier from FE1 to FE2 decreases dramatically, which suggests that FE switching under the electric field can be easily achieved~\cite{M1M1P2X6.SB}.

The $\alpha$- and $\beta$-Bi$_4X_4$ ($X$=Br, I) crystal structures are shown in Fig.~\ref{fig:S2}(a). Among them, the $\alpha$-Bi$_4$Br$_4$ crystallizes in the monoclinic SG $C2/m$ (\#12), where the two adjacent layers are related by $C_{2z}$ rotation~\cite{BiX}. The Bi$_4$Br$_4$ monolayer, a large-gap quantum spin Hall insulator~\cite{BiBr-1layer.nl}, hosts IS and lacks polarization. In order to introduce ferroelectricity, we construct the $\pi$-bilayer structure.
The two FE phases with SG $Cm$ (\#8) and the PE phase with SG $Pc$ (\#7) are obtained in our calculations [FE1: $\textbf{v}_{||}=(0.225,0)$; FE2: $\textbf{v}_{||}=(0.275,0.5)$; PE: $\textbf{v}_{||}=(0.25,0.25)$], whose structures are shown in Fig.~\ref{fig:S2}(b).
The two FE structures linked by $m_z$ symmetry have the opposite vertical polarizations (0.05 pC/m), and are dynamically stable, which is confirmed by phonon spectra in Fig.~\ref{fig:S2}(c). A glide symmetry $g_z$ in the PE phase forbids vertical polarization.
From the total energy as a function of the $\textbf{v}_{||}$ in Fig.~\ref{fig:S2}(e), we obtained an FE switching pathway realized by interlayer sliding as depicted by the dashed line.
The typical energy double well structure is clearly visible in Fig.~\ref{fig:S2}(g), with a low energy barrier of 20.8 meV per unit cell.
In addition, the Bi$_4$I$_4$ $\pi$-bilayer with vertical polarizations of 0.24 pC/m has the same FE behaviors [FE1: $\textbf{v}_{||}=(0.23,0)$; FE2: $\textbf{v}_{||}=(0.27,0.5)$; PE: $\textbf{v}_{||}=(0.25,0.25)$], as shown in Figs.~\ref{fig:S2}(d), \ref{fig:S2}(f), and \ref{fig:S2}(g).

\begin{figure}[!htb]
\centering
\includegraphics[width=8.5 cm]{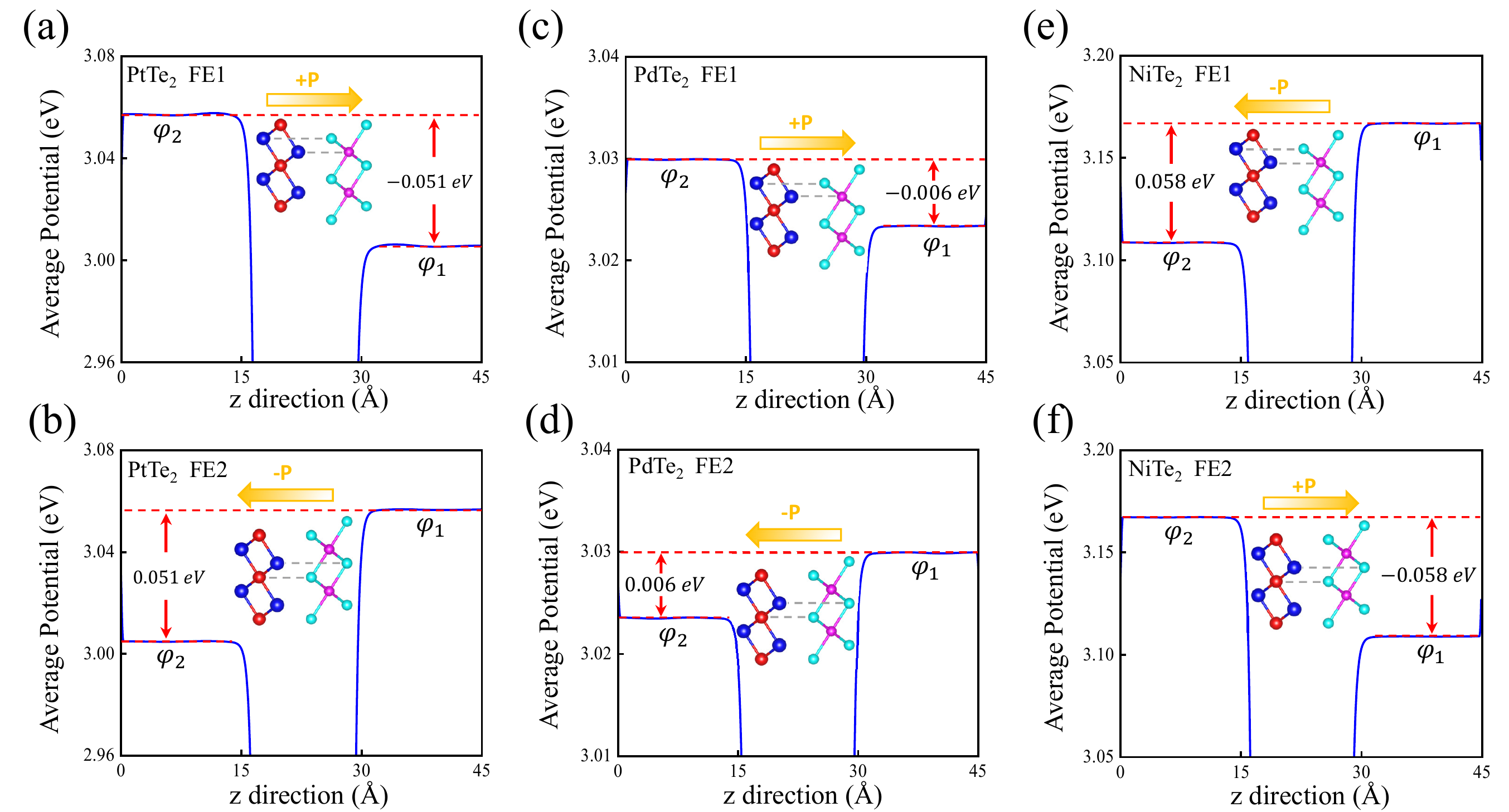}
\caption{(Color online)
Plane-averaged electrostatic potentials along the $z$ direction of two FE phases (FE1 and FE2) for (a, b) PtTe$_2$, (c, d) PdTe$_2$, and (e, f) NiTe$_2$ $\pi$ bilayers. Inset presents the corresponding atomic structure. The orange arrow indicates polarization direction.
} \label{fig:S3}
\end{figure}

\begin{figure}[!htb]
\centering
\includegraphics[width=8.5 cm]{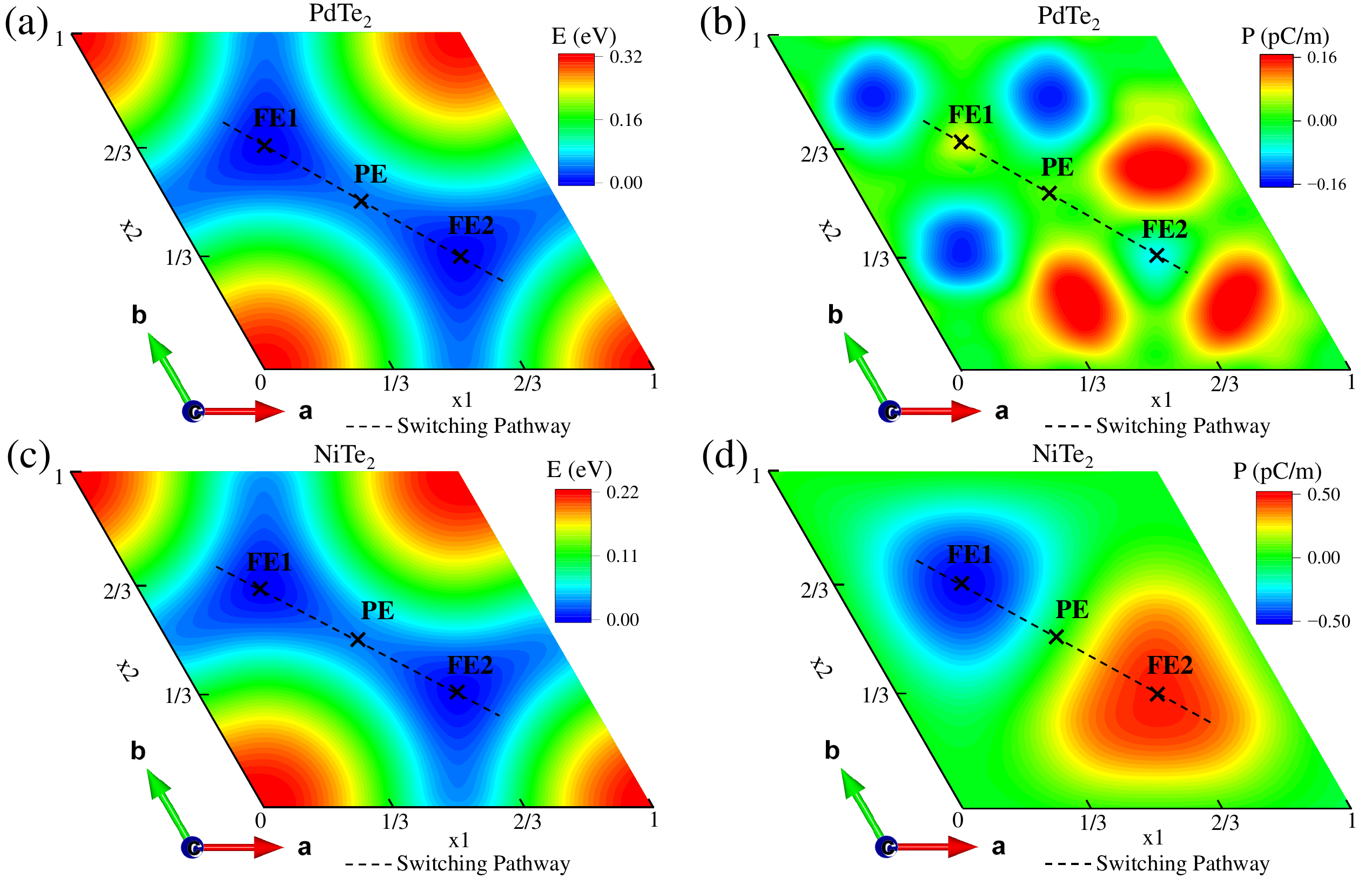}
\caption{(Color online)
Contour plots of total energy and vertical polarization versus the $\textbf{v}_{||}$ for (a, b) PdTe$_2$ and (c, d) NiTe$_2$ $\pi$ bilayers. The total energies corresponding to two FE phases (FE1 and FE2) are set to zero. FE switching pathways are marked with black dotted lines.
} \label{fig:S4}
\end{figure}

\section{FERROELECTRIC IN $M$Te$_2$ $\pi$ BILAYERS}
\label{SM3}
Due to the presence of spontaneous vertical polarization, there is a built-in electric field, resulting in different vacuum levels on the upper and lower sides of the FE slab. As shown by plane-averaged electrostatic potential along the $z$ direction in Fig.~\ref{fig:S3}, the discontinuity of vacuum levels $(\Delta\varphi=\varphi_1 - \varphi_2)$ is obvious, where $\varphi_1$ and $\varphi_2$ represent the vacuum levels of the top and bottom sides, respectively. For PtTe$_2$ $\pi$ bilayers in Figs.~\ref{fig:S3}(a) and \ref{fig:S3}(b), the negative $\Delta\varphi$ (-0.051 eV) of the FE1 phase indicates an upward vertical polarization, while a vertical polarization of the FE2 phase with equal magnitude and the opposite direction is demonstrated by the positive $\Delta\varphi$ (0.051 eV) with equal magnitude. The same is true for PdTe$_2$ and NiTe$_2$ $\pi$ bilayers, as shown in Figs.~\ref{fig:S3}(c)-\ref{fig:S3}(f). The larger the $\Delta\varphi$, the greater the spontaneous polarization, which is also reflected in $M$Te$_2$ FE $\pi$ bilayers. Contour plots of total energy and vertical polarization versus the $\textbf{v}_{||}$ for PdTe$_2$ and NiTe$_2$ $\pi$ bilayers are shown in Fig.~\ref{fig:S4}, which are similar to those of the PtTe$_2$ $\pi$ bilayer.

\begin{figure}[!htb]
\centering
\includegraphics[width=8.5 cm]{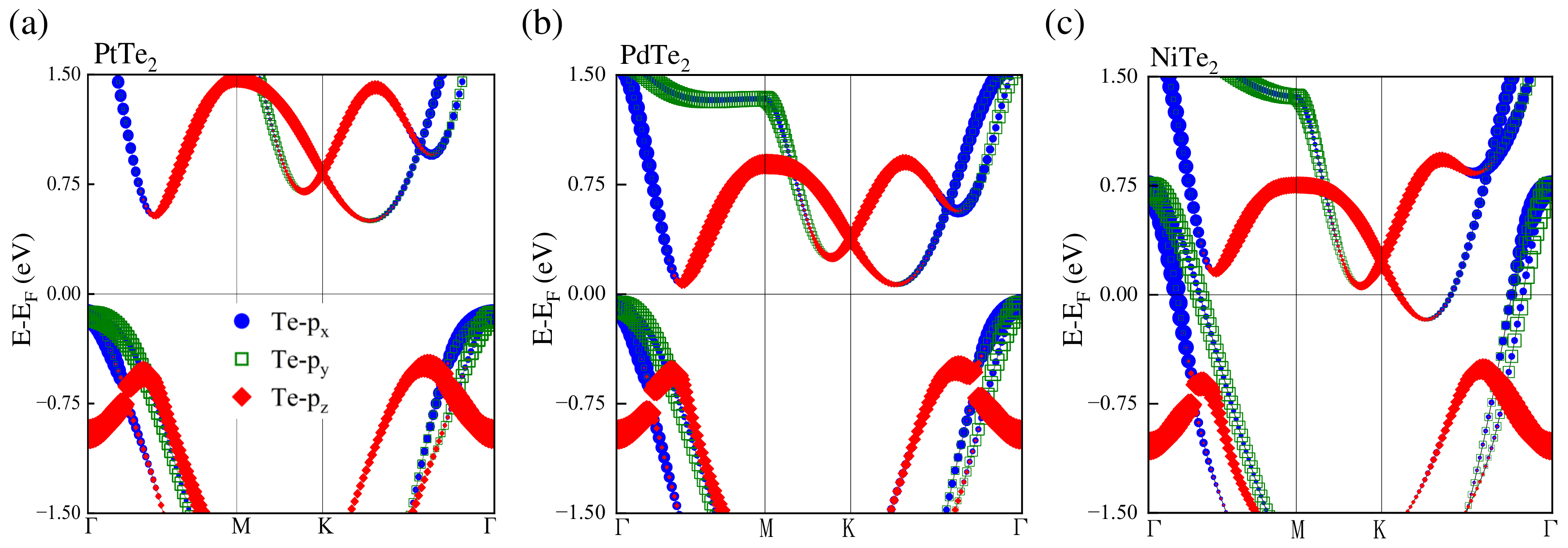}
\caption{(Color online)
Orbital-resolved band structures for (a) PtTe$_2$, (b) PdTe$_2$, and (c) NiTe$_2$ 1T monolayers. The sizes of the blue circles, green hollow squares, and red diamonds represent the weights of Te-$p_x$, Te-$p_y$, and Te-$p_z$ orbitals, respectively.
} \label{fig:S5}
\end{figure}

\begin{figure}[!htb]
\centering
\includegraphics[width=8.5 cm]{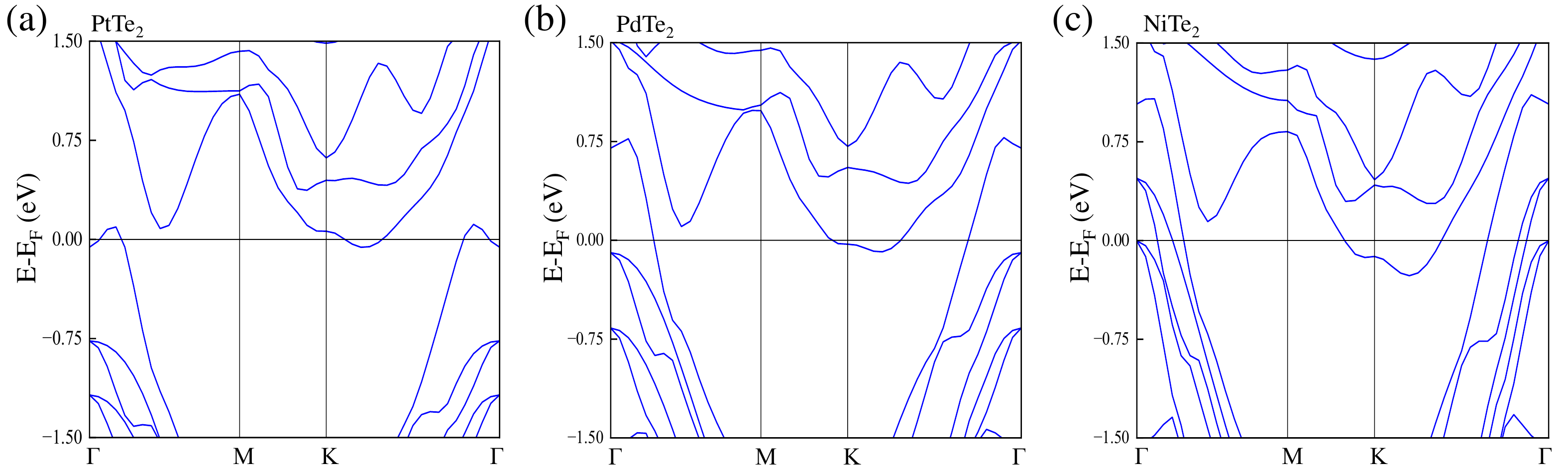}
\caption{(Color online)
Band structures with the HSE06 functional for (a) PtTe$_2$, (b) PdTe$_2$, and (c) NiTe$_2$ FE $\pi$ bilayers.
} \label{fig:S6}
\end{figure}

\begin{figure*}[!htb]
\centering
\includegraphics[width=14.5 cm]{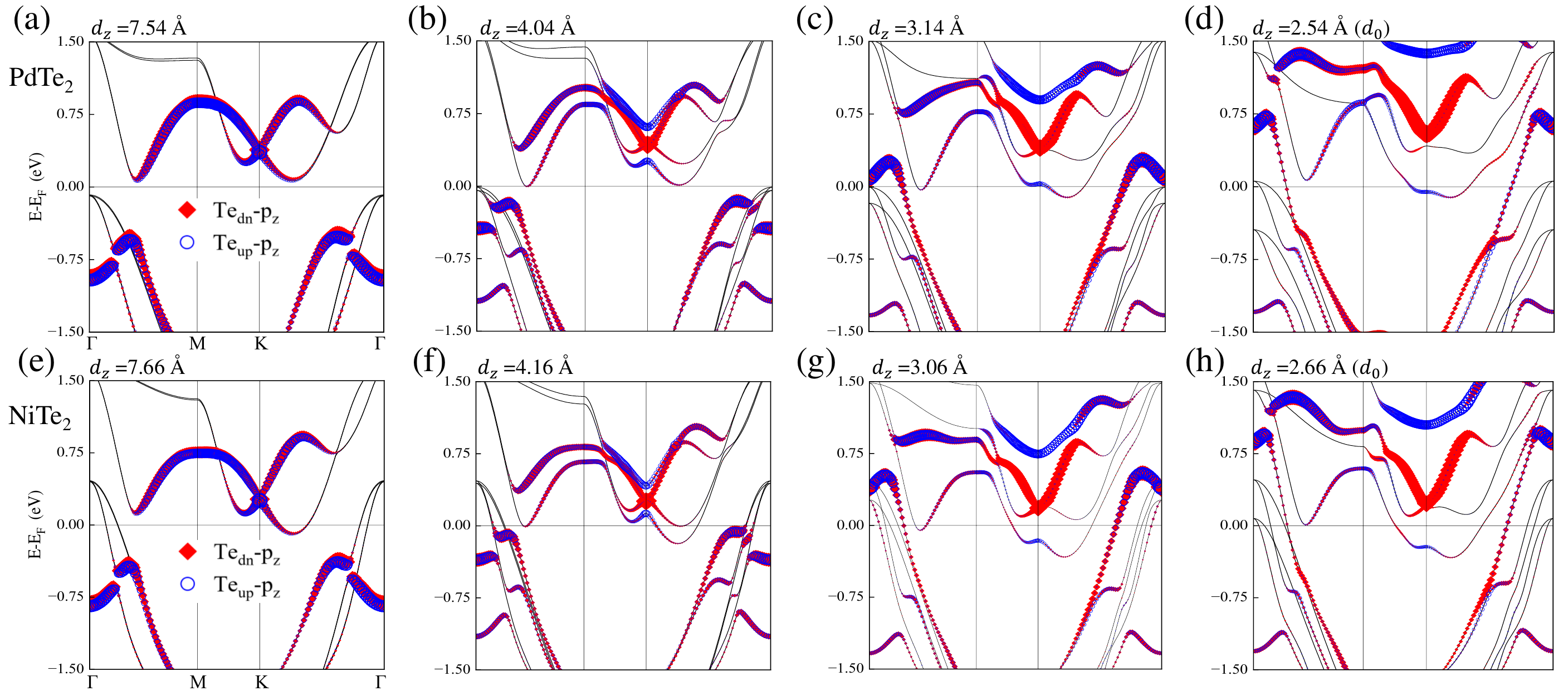}
\caption{(Color online)
Evolution of orbital-resolved band structures with interlayer distance $d_z$ for FE $\pi$ bilayers. For the PdTe$_2$ $\pi$ bilayer, (a) $d_z$=7.54 \AA, (b) $d_z$=4.04 \AA, (c) $d_z$=3.14 \AA, and (d) $d_z$=2.54 \AA ($d_0$). For the NiTe$_2$ $\pi$ bilayer, (e) $d_z$=7.66 \AA, (f) $d_z$=4.16 \AA, (g) $d_z$=3.06 \AA, and (h) $d_z$=2.66 \AA ($d_0$). The definition of $d_z$ is shown in the inset of Fig.~\ref{fig:3}(b) and $d_0$ represents the actual $d_z$ obtained by full relaxation. The sizes of the red diamonds and blue hollow circles represent the weights of Te$_{dn}$-$p_z$ and Te$_{up}$-$p_z$ orbitals, respectively.
} \label{fig:S7}
\end{figure*}

\section{BAND STRUCTURE OF $M$Te$_2$}
\label{SM4}
Orbital-resolved band structures for $M$Te$_2$ 1T monolayers are shown in Fig.~\ref{fig:S5}. In order to check the metallicity of $M$Te$_2$ FE $\pi$ bilayers, we employ the HSE06 functional to obtain more accurate band structures. As shown in Fig.~\ref{fig:S6}, the bilayers are metallic, which is consistent with the results of the PBE functional [see Figs.~\ref{fig:3}(f), \ref{fig:S7}(d), and \ref{fig:S7}(h)]. The difference in band structure obtained by the HSE06 function and the PBE function is negligible.

\section{ORIGIN OF METALLICITY FOR PdTe$_2$ AND NiTe$_2$ $\pi$ BILAYERS}
\label{SM5}
As with the PtTe$_2$ FE $\pi$ bilayer described in the main text, interlayer vdW charge transfer also causes the valence band rise to introduce metallicity into PdTe$_2$ and NiTe$_2$ FE $\pi$ bilayers. With the increase of the interlayer vdW charge transfer due to the reduced $d_z$ in Fig.~\ref{fig:S7}(a)-\ref{fig:S7}(d), the valence band dominated by the Te$_{up}$-$p_z$ and Te$_{dn}$-$p_z$ orbitals gradually rises above $E_F$ near the $\Gamma$ point, which transforms the PdTe$_2$ FE $\pi$ bilayer from an insulator to a metal. For the NiTe$_2$ 1T monolayer in Fig.~\ref{fig:S5}(c), although there is a non-zero density of states near $E_F$, only two isolated valence bands dominated by the Te-$p_x$ and Te-$p_y$ orbitals pass through $E_F$, limiting conductivity. As shown in Figs.~\ref{fig:S7}(e)-\ref{fig:S7}(h), the same valence band rise gives the NiTe$_2$ FE $\pi$ bilayer more conductive electron states.

\newpage

\begin{thebibliography}{60}%
\makeatletter
\providecommand \@ifxundefined [1]{%
 \@ifx{#1\undefined}
}%
\providecommand \@ifnum [1]{%
 \ifnum #1\expandafter \@firstoftwo
 \else \expandafter \@secondoftwo
 \fi
}%
\providecommand \@ifx [1]{%
 \ifx #1\expandafter \@firstoftwo
 \else \expandafter \@secondoftwo
 \fi
}%
\providecommand \natexlab [1]{#1}%
\providecommand \enquote  [1]{``#1''}%
\providecommand \bibnamefont  [1]{#1}%
\providecommand \bibfnamefont [1]{#1}%
\providecommand \citenamefont [1]{#1}%
\providecommand \href@noop [0]{\@secondoftwo}%
\providecommand \href [0]{\begingroup \@sanitize@url \@href}%
\providecommand \@href[1]{\@@startlink{#1}\@@href}%
\providecommand \@@href[1]{\endgroup#1\@@endlink}%
\providecommand \@sanitize@url [0]{\catcode `\\12\catcode `\$12\catcode
  `\&12\catcode `\#12\catcode `\^12\catcode `\_12\catcode `\%12\relax}%
\providecommand \@@startlink[1]{}%
\providecommand \@@endlink[0]{}%
\providecommand \url  [0]{\begingroup\@sanitize@url \@url }%
\providecommand \@url [1]{\endgroup\@href {#1}{\urlprefix }}%
\providecommand \urlprefix  [0]{URL }%
\providecommand \Eprint [0]{\href }%
\providecommand \doibase [0]{http://dx.doi.org/}%
\providecommand \selectlanguage [0]{\@gobble}%
\providecommand \bibinfo  [0]{\@secondoftwo}%
\providecommand \bibfield  [0]{\@secondoftwo}%
\providecommand \translation [1]{[#1]}%
\providecommand \BibitemOpen [0]{}%
\providecommand \bibitemStop [0]{}%
\providecommand \bibitemNoStop [0]{.\EOS\space}%
\providecommand \EOS [0]{\spacefactor3000\relax}%
\providecommand \BibitemShut  [1]{\csname bibitem#1\endcsname}%
\let\auto@bib@innerbib\@empty
\bibitem [{\citenamefont {Valasek}(1921)}]{PhysRev.17.475}%
  \BibitemOpen
  \bibfield  {author} {\bibinfo {author} {\bibfnamefont {J.}~\bibnamefont
  {Valasek}},\ }\href {\doibase 10.1103/PhysRev.17.475} {\bibfield  {journal}
  {\bibinfo  {journal} {Phys. Rev.}\ }\textbf {\bibinfo {volume} {17}},\
  \bibinfo {pages} {475} (\bibinfo {year} {1921})}\BibitemShut {NoStop}%
\bibitem [{\citenamefont {Shen}\ \emph {et~al.}(2019)\citenamefont {Shen},
  \citenamefont {Fang}, \citenamefont {Tian},\ and\ \citenamefont
  {Duan}}]{ACSaelm.2019}%
  \BibitemOpen
  \bibfield  {author} {\bibinfo {author} {\bibfnamefont {X.-W.}\ \bibnamefont
  {Shen}}, \bibinfo {author} {\bibfnamefont {Y.-W.}\ \bibnamefont {Fang}},
  \bibinfo {author} {\bibfnamefont {B.-B.}\ \bibnamefont {Tian}}, \ and\
  \bibinfo {author} {\bibfnamefont {C.-G.}\ \bibnamefont {Duan}},\ }\href
  {\doibase 10.1021/acsaelm.9b00146} {\bibfield  {journal} {\bibinfo  {journal}
  {ACS Applied Electronic Materials}\ }\textbf {\bibinfo {volume} {1}},\
  \bibinfo {pages} {1133} (\bibinfo {year} {2019})}\BibitemShut {NoStop}%
\bibitem [{\citenamefont {Yuan}\ \emph {et~al.}(2019)\citenamefont {Yuan},
  \citenamefont {Luo}, \citenamefont {Chan}, \citenamefont {Xiao},
  \citenamefont {Dai}, \citenamefont {Xie},\ and\ \citenamefont
  {Hao}}]{d1T-MoTe2.NC}%
  \BibitemOpen
  \bibfield  {author} {\bibinfo {author} {\bibfnamefont {S.}~\bibnamefont
  {Yuan}}, \bibinfo {author} {\bibfnamefont {X.}~\bibnamefont {Luo}}, \bibinfo
  {author} {\bibfnamefont {H.~L.}\ \bibnamefont {Chan}}, \bibinfo {author}
  {\bibfnamefont {C.}~\bibnamefont {Xiao}}, \bibinfo {author} {\bibfnamefont
  {Y.}~\bibnamefont {Dai}}, \bibinfo {author} {\bibfnamefont {M.}~\bibnamefont
  {Xie}}, \ and\ \bibinfo {author} {\bibfnamefont {J.}~\bibnamefont {Hao}},\
  }\href {\doibase 10.1038/s41467-019-09669-x} {\bibfield  {journal} {\bibinfo
  {journal} {Nature Communications}\ }\textbf {\bibinfo {volume} {10}},\
  \bibinfo {pages} {1775} (\bibinfo {year} {2019})}\BibitemShut {NoStop}%
\bibitem [{\citenamefont {Ding}\ \emph {et~al.}(2017)\citenamefont {Ding},
  \citenamefont {Zhu}, \citenamefont {Wang}, \citenamefont {Gao}, \citenamefont
  {Xiao}, \citenamefont {Gu}, \citenamefont {Zhang},\ and\ \citenamefont
  {Zhu}}]{In2Se3.NC}%
  \BibitemOpen
  \bibfield  {author} {\bibinfo {author} {\bibfnamefont {W.}~\bibnamefont
  {Ding}}, \bibinfo {author} {\bibfnamefont {J.}~\bibnamefont {Zhu}}, \bibinfo
  {author} {\bibfnamefont {Z.}~\bibnamefont {Wang}}, \bibinfo {author}
  {\bibfnamefont {Y.}~\bibnamefont {Gao}}, \bibinfo {author} {\bibfnamefont
  {D.}~\bibnamefont {Xiao}}, \bibinfo {author} {\bibfnamefont {Y.}~\bibnamefont
  {Gu}}, \bibinfo {author} {\bibfnamefont {Z.}~\bibnamefont {Zhang}}, \ and\
  \bibinfo {author} {\bibfnamefont {W.}~\bibnamefont {Zhu}},\ }\href {\doibase
  10.1038/ncomms14956} {\bibfield  {journal} {\bibinfo  {journal} {Nature
  Communications}\ }\textbf {\bibinfo {volume} {8}},\ \bibinfo {pages} {14956}
  (\bibinfo {year} {2017})}\BibitemShut {NoStop}%
\bibitem [{\citenamefont {Fei}\ \emph {et~al.}(2016)\citenamefont {Fei},
  \citenamefont {Kang},\ and\ \citenamefont {Yang}}]{SnSe.PRL}%
  \BibitemOpen
  \bibfield  {author} {\bibinfo {author} {\bibfnamefont {R.}~\bibnamefont
  {Fei}}, \bibinfo {author} {\bibfnamefont {W.}~\bibnamefont {Kang}}, \ and\
  \bibinfo {author} {\bibfnamefont {L.}~\bibnamefont {Yang}},\ }\href {\doibase
  10.1103/PhysRevLett.117.097601} {\bibfield  {journal} {\bibinfo  {journal}
  {Phys. Rev. Lett.}\ }\textbf {\bibinfo {volume} {117}},\ \bibinfo {pages}
  {097601} (\bibinfo {year} {2016})}\BibitemShut {NoStop}%
\bibitem [{\citenamefont {Qi}\ \emph {et~al.}(2021)\citenamefont {Qi},
  \citenamefont {Ruan},\ and\ \citenamefont {Zeng}}]{AdMa.2021}%
  \BibitemOpen
  \bibfield  {author} {\bibinfo {author} {\bibfnamefont {L.}~\bibnamefont
  {Qi}}, \bibinfo {author} {\bibfnamefont {S.}~\bibnamefont {Ruan}}, \ and\
  \bibinfo {author} {\bibfnamefont {Y.-J.}\ \bibnamefont {Zeng}},\ }\href
  {\doibase https://doi.org/10.1002/adma.202005098} {\bibfield  {journal}
  {\bibinfo  {journal} {Advanced Materials}\ }\textbf {\bibinfo {volume}
  {33}},\ \bibinfo {pages} {2005098} (\bibinfo {year} {2021})}\BibitemShut
  {NoStop}%
\bibitem [{\citenamefont {Chanana}\ and\ \citenamefont
  {Waghmare}(2019)}]{NbN.PRL}%
  \BibitemOpen
  \bibfield  {author} {\bibinfo {author} {\bibfnamefont {A.}~\bibnamefont
  {Chanana}}\ and\ \bibinfo {author} {\bibfnamefont {U.~V.}\ \bibnamefont
  {Waghmare}},\ }\href {\doibase 10.1103/PhysRevLett.123.037601} {\bibfield
  {journal} {\bibinfo  {journal} {Phys. Rev. Lett.}\ }\textbf {\bibinfo
  {volume} {123}},\ \bibinfo {pages} {037601} (\bibinfo {year}
  {2019})}\BibitemShut {NoStop}%
\bibitem [{\citenamefont {Manzeli}\ \emph {et~al.}(2017)\citenamefont
  {Manzeli}, \citenamefont {Ovchinnikov}, \citenamefont {Pasquier},
  \citenamefont {Yazyev},\ and\ \citenamefont {Kis}}]{TMD.NRM}%
  \BibitemOpen
  \bibfield  {author} {\bibinfo {author} {\bibfnamefont {S.}~\bibnamefont
  {Manzeli}}, \bibinfo {author} {\bibfnamefont {D.}~\bibnamefont
  {Ovchinnikov}}, \bibinfo {author} {\bibfnamefont {D.}~\bibnamefont
  {Pasquier}}, \bibinfo {author} {\bibfnamefont {O.~V.}\ \bibnamefont
  {Yazyev}}, \ and\ \bibinfo {author} {\bibfnamefont {A.}~\bibnamefont {Kis}},\
  }\href {\doibase 10.1038/natrevmats.2017.33} {\bibfield  {journal} {\bibinfo
  {journal} {Nature Reviews Materials}\ }\textbf {\bibinfo {volume} {2}},\
  \bibinfo {pages} {17033} (\bibinfo {year} {2017})}\BibitemShut {NoStop}%
\bibitem [{\citenamefont {Yin}\ \emph {et~al.}(2021)\citenamefont {Yin},
  \citenamefont {Tang}, \citenamefont {Zheng}, \citenamefont {Gao},
  \citenamefont {Wu}, \citenamefont {Zhang}, \citenamefont {Chhowalla},
  \citenamefont {Chen},\ and\ \citenamefont {Wee}}]{TMD.CSR}%
  \BibitemOpen
  \bibfield  {author} {\bibinfo {author} {\bibfnamefont {X.}~\bibnamefont
  {Yin}}, \bibinfo {author} {\bibfnamefont {C.~S.}\ \bibnamefont {Tang}},
  \bibinfo {author} {\bibfnamefont {Y.}~\bibnamefont {Zheng}}, \bibinfo
  {author} {\bibfnamefont {J.}~\bibnamefont {Gao}}, \bibinfo {author}
  {\bibfnamefont {J.}~\bibnamefont {Wu}}, \bibinfo {author} {\bibfnamefont
  {H.}~\bibnamefont {Zhang}}, \bibinfo {author} {\bibfnamefont
  {M.}~\bibnamefont {Chhowalla}}, \bibinfo {author} {\bibfnamefont
  {W.}~\bibnamefont {Chen}}, \ and\ \bibinfo {author} {\bibfnamefont
  {A.~T.~S.}\ \bibnamefont {Wee}},\ }\href {\doibase 10.1039/D1CS00236H}
  {\bibfield  {journal} {\bibinfo  {journal} {Chem. Soc. Rev.}\ }\textbf
  {\bibinfo {volume} {50}},\ \bibinfo {pages} {10087} (\bibinfo {year}
  {2021})}\BibitemShut {NoStop}%
\bibitem [{\citenamefont {Novoselov}\ \emph {et~al.}(2004)\citenamefont
  {Novoselov}, \citenamefont {Geim}, \citenamefont {Morozov}, \citenamefont
  {Jiang}, \citenamefont {Zhang}, \citenamefont {Dubonos}, \citenamefont
  {Grigorieva},\ and\ \citenamefont {Firsov}}]{graphene.Science}%
  \BibitemOpen
  \bibfield  {author} {\bibinfo {author} {\bibfnamefont {K.~S.}\ \bibnamefont
  {Novoselov}}, \bibinfo {author} {\bibfnamefont {A.~K.}\ \bibnamefont {Geim}},
  \bibinfo {author} {\bibfnamefont {S.~V.}\ \bibnamefont {Morozov}}, \bibinfo
  {author} {\bibfnamefont {D.}~\bibnamefont {Jiang}}, \bibinfo {author}
  {\bibfnamefont {Y.}~\bibnamefont {Zhang}}, \bibinfo {author} {\bibfnamefont
  {S.~V.}\ \bibnamefont {Dubonos}}, \bibinfo {author} {\bibfnamefont {I.~V.}\
  \bibnamefont {Grigorieva}}, \ and\ \bibinfo {author} {\bibfnamefont {A.~A.}\
  \bibnamefont {Firsov}},\ }\href {\doibase 10.1126/science.1102896} {\bibfield
   {journal} {\bibinfo  {journal} {Science}\ }\textbf {\bibinfo {volume}
  {306}},\ \bibinfo {pages} {666} (\bibinfo {year} {2004})}\BibitemShut
  {NoStop}%
\bibitem [{\citenamefont {Tavakoli}\ \emph {et~al.}(2021)\citenamefont
  {Tavakoli}, \citenamefont {Park}, \citenamefont {Mwaura}, \citenamefont
  {Saravanapavanantham}, \citenamefont {Bulović},\ and\ \citenamefont
  {Kong}}]{BN.AFM}%
  \BibitemOpen
  \bibfield  {author} {\bibinfo {author} {\bibfnamefont {M.~M.}\ \bibnamefont
  {Tavakoli}}, \bibinfo {author} {\bibfnamefont {J.-H.}\ \bibnamefont {Park}},
  \bibinfo {author} {\bibfnamefont {J.}~\bibnamefont {Mwaura}}, \bibinfo
  {author} {\bibfnamefont {M.}~\bibnamefont {Saravanapavanantham}}, \bibinfo
  {author} {\bibfnamefont {V.}~\bibnamefont {Bulović}}, \ and\ \bibinfo
  {author} {\bibfnamefont {J.}~\bibnamefont {Kong}},\ }\href {\doibase
  https://doi.org/10.1002/adfm.202101238} {\bibfield  {journal} {\bibinfo
  {journal} {Advanced Functional Materials}\ }\textbf {\bibinfo {volume}
  {31}},\ \bibinfo {pages} {2101238} (\bibinfo {year} {2021})}\BibitemShut
  {NoStop}%
\bibitem [{\citenamefont {Lei}\ and\ \citenamefont
  {Menghao}(2017)}]{Nano.2017}%
  \BibitemOpen
  \bibfield  {author} {\bibinfo {author} {\bibfnamefont {L.}~\bibnamefont
  {Lei}}\ and\ \bibinfo {author} {\bibfnamefont {W.}~\bibnamefont {Menghao}},\
  }\href {\doibase 10.1021/acsnano.7b02756} {\bibfield  {journal} {\bibinfo
  {journal} {ACS Nano}\ }\textbf {\bibinfo {volume} {11}},\ \bibinfo {pages}
  {6382–6388} (\bibinfo {year} {2017})}\BibitemShut {NoStop}%
\bibitem [{\citenamefont {Zhong}\ \emph {et~al.}(2021)\citenamefont {Zhong},
  \citenamefont {Ren}, \citenamefont {Zhang}, \citenamefont {Gao},\ and\
  \citenamefont {Wu}}]{MoA2N4.JMCA}%
  \BibitemOpen
  \bibfield  {author} {\bibinfo {author} {\bibfnamefont {T.}~\bibnamefont
  {Zhong}}, \bibinfo {author} {\bibfnamefont {Y.}~\bibnamefont {Ren}}, \bibinfo
  {author} {\bibfnamefont {Z.}~\bibnamefont {Zhang}}, \bibinfo {author}
  {\bibfnamefont {J.}~\bibnamefont {Gao}}, \ and\ \bibinfo {author}
  {\bibfnamefont {M.}~\bibnamefont {Wu}},\ }\href {\doibase 10.1039/D1TA02645C}
  {\bibfield  {journal} {\bibinfo  {journal} {J. Mater. Chem. A}\ }\textbf
  {\bibinfo {volume} {9}},\ \bibinfo {pages} {19659} (\bibinfo {year}
  {2021})}\BibitemShut {NoStop}%
\bibitem [{\citenamefont {Xiao}\ \emph {et~al.}(2022)\citenamefont {Xiao},
  \citenamefont {Gao}, \citenamefont {Jiang}, \citenamefont {Gan},
  \citenamefont {Zhang},\ and\ \citenamefont {Li}}]{npj.8.138}%
  \BibitemOpen
  \bibfield  {author} {\bibinfo {author} {\bibfnamefont {R.-C.}\ \bibnamefont
  {Xiao}}, \bibinfo {author} {\bibfnamefont {Y.}~\bibnamefont {Gao}}, \bibinfo
  {author} {\bibfnamefont {H.}~\bibnamefont {Jiang}}, \bibinfo {author}
  {\bibfnamefont {W.}~\bibnamefont {Gan}}, \bibinfo {author} {\bibfnamefont
  {C.}~\bibnamefont {Zhang}}, \ and\ \bibinfo {author} {\bibfnamefont
  {H.}~\bibnamefont {Li}},\ }\href {\doibase 10.1038/s41524-022-00828-1}
  {\bibfield  {journal} {\bibinfo  {journal} {npj Computational Materials}\
  }\textbf {\bibinfo {volume} {8}},\ \bibinfo {pages} {138} (\bibinfo {year}
  {2022})}\BibitemShut {NoStop}%
\bibitem [{\citenamefont {Sun}\ \emph {et~al.}(2022)\citenamefont {Sun},
  \citenamefont {Wang}, \citenamefont {Li}, \citenamefont {Li}, \citenamefont
  {Yu}, \citenamefont {Bai}, \citenamefont {Zhang},\ and\ \citenamefont
  {Cheng}}]{LaBr2.npj}%
  \BibitemOpen
  \bibfield  {author} {\bibinfo {author} {\bibfnamefont {W.}~\bibnamefont
  {Sun}}, \bibinfo {author} {\bibfnamefont {W.}~\bibnamefont {Wang}}, \bibinfo
  {author} {\bibfnamefont {H.}~\bibnamefont {Li}}, \bibinfo {author}
  {\bibfnamefont {X.}~\bibnamefont {Li}}, \bibinfo {author} {\bibfnamefont
  {Z.}~\bibnamefont {Yu}}, \bibinfo {author} {\bibfnamefont {Y.}~\bibnamefont
  {Bai}}, \bibinfo {author} {\bibfnamefont {G.}~\bibnamefont {Zhang}}, \ and\
  \bibinfo {author} {\bibfnamefont {Z.}~\bibnamefont {Cheng}},\ }\href
  {\doibase 10.1038/s41524-022-00833-4} {\bibfield  {journal} {\bibinfo
  {journal} {npj Computational Materials}\ }\textbf {\bibinfo {volume} {8}},\
  \bibinfo {pages} {159} (\bibinfo {year} {2022})}\BibitemShut {NoStop}%
\bibitem [{\citenamefont {Zhang}\ \emph {et~al.}(2022)\citenamefont {Zhang},
  \citenamefont {Xu}, \citenamefont {Huang}, \citenamefont {Dai},\ and\
  \citenamefont {Ma}}]{FeCl2.npj}%
  \BibitemOpen
  \bibfield  {author} {\bibinfo {author} {\bibfnamefont {T.}~\bibnamefont
  {Zhang}}, \bibinfo {author} {\bibfnamefont {X.}~\bibnamefont {Xu}}, \bibinfo
  {author} {\bibfnamefont {B.}~\bibnamefont {Huang}}, \bibinfo {author}
  {\bibfnamefont {Y.}~\bibnamefont {Dai}}, \ and\ \bibinfo {author}
  {\bibfnamefont {Y.}~\bibnamefont {Ma}},\ }\href {\doibase
  10.1038/s41524-022-00748-0} {\bibfield  {journal} {\bibinfo  {journal} {npj
  Computational Materials}\ }\textbf {\bibinfo {volume} {8}},\ \bibinfo {pages}
  {64} (\bibinfo {year} {2022})}\BibitemShut {NoStop}%
\bibitem [{\citenamefont {Liu}\ \emph {et~al.}(2020)\citenamefont {Liu},
  \citenamefont {Pyatakov},\ and\ \citenamefont {Ren}}]{VS2.PRL}%
  \BibitemOpen
  \bibfield  {author} {\bibinfo {author} {\bibfnamefont {X.}~\bibnamefont
  {Liu}}, \bibinfo {author} {\bibfnamefont {A.~P.}\ \bibnamefont {Pyatakov}}, \
  and\ \bibinfo {author} {\bibfnamefont {W.}~\bibnamefont {Ren}},\ }\href
  {\doibase 10.1103/PhysRevLett.125.247601} {\bibfield  {journal} {\bibinfo
  {journal} {Phys. Rev. Lett.}\ }\textbf {\bibinfo {volume} {125}},\ \bibinfo
  {pages} {247601} (\bibinfo {year} {2020})}\BibitemShut {NoStop}%
\bibitem [{\citenamefont {Yasuda}\ \emph {et~al.}(2021)\citenamefont {Yasuda},
  \citenamefont {Wang}, \citenamefont {Watanabe}, \citenamefont {Taniguchi},\
  and\ \citenamefont {Jarillo-Herrero}}]{BN.Science1}%
  \BibitemOpen
  \bibfield  {author} {\bibinfo {author} {\bibfnamefont {K.}~\bibnamefont
  {Yasuda}}, \bibinfo {author} {\bibfnamefont {X.}~\bibnamefont {Wang}},
  \bibinfo {author} {\bibfnamefont {K.}~\bibnamefont {Watanabe}}, \bibinfo
  {author} {\bibfnamefont {T.}~\bibnamefont {Taniguchi}}, \ and\ \bibinfo
  {author} {\bibfnamefont {P.}~\bibnamefont {Jarillo-Herrero}},\ }\href
  {\doibase 10.1126/science.abd3230} {\bibfield  {journal} {\bibinfo  {journal}
  {Science}\ }\textbf {\bibinfo {volume} {372}},\ \bibinfo {pages} {1458}
  (\bibinfo {year} {2021})}\BibitemShut {NoStop}%
\bibitem [{\citenamefont {Stern}\ \emph {et~al.}(2021)\citenamefont {Stern},
  \citenamefont {Waschitz}, \citenamefont {Cao}, \citenamefont {Nevo},
  \citenamefont {Watanabe}, \citenamefont {Taniguchi}, \citenamefont {Sela},
  \citenamefont {Urbakh}, \citenamefont {Hod},\ and\ \citenamefont
  {Shalom}}]{BN.Science2}%
  \BibitemOpen
  \bibfield  {author} {\bibinfo {author} {\bibfnamefont {M.~V.}\ \bibnamefont
  {Stern}}, \bibinfo {author} {\bibfnamefont {Y.}~\bibnamefont {Waschitz}},
  \bibinfo {author} {\bibfnamefont {W.}~\bibnamefont {Cao}}, \bibinfo {author}
  {\bibfnamefont {I.}~\bibnamefont {Nevo}}, \bibinfo {author} {\bibfnamefont
  {K.}~\bibnamefont {Watanabe}}, \bibinfo {author} {\bibfnamefont
  {T.}~\bibnamefont {Taniguchi}}, \bibinfo {author} {\bibfnamefont
  {E.}~\bibnamefont {Sela}}, \bibinfo {author} {\bibfnamefont {M.}~\bibnamefont
  {Urbakh}}, \bibinfo {author} {\bibfnamefont {O.}~\bibnamefont {Hod}}, \ and\
  \bibinfo {author} {\bibfnamefont {M.~B.}\ \bibnamefont {Shalom}},\ }\href
  {\doibase 10.1126/science.abe8177} {\bibfield  {journal} {\bibinfo  {journal}
  {Science}\ }\textbf {\bibinfo {volume} {372}},\ \bibinfo {pages} {1462}
  (\bibinfo {year} {2021})}\BibitemShut {NoStop}%
\bibitem [{\citenamefont {Wang}\ \emph {et~al.}(2022)\citenamefont {Wang},
  \citenamefont {Yasuda}, \citenamefont {Zhang}, \citenamefont {Liu},
  \citenamefont {Watanabe}, \citenamefont {Taniguchi}, \citenamefont {Hone},
  \citenamefont {Fu},\ and\ \citenamefont {Jarillo-Herrero}}]{H-MX2.NatNan}%
  \BibitemOpen
  \bibfield  {author} {\bibinfo {author} {\bibfnamefont {X.}~\bibnamefont
  {Wang}}, \bibinfo {author} {\bibfnamefont {K.}~\bibnamefont {Yasuda}},
  \bibinfo {author} {\bibfnamefont {Y.}~\bibnamefont {Zhang}}, \bibinfo
  {author} {\bibfnamefont {S.}~\bibnamefont {Liu}}, \bibinfo {author}
  {\bibfnamefont {K.}~\bibnamefont {Watanabe}}, \bibinfo {author}
  {\bibfnamefont {T.}~\bibnamefont {Taniguchi}}, \bibinfo {author}
  {\bibfnamefont {J.}~\bibnamefont {Hone}}, \bibinfo {author} {\bibfnamefont
  {L.}~\bibnamefont {Fu}}, \ and\ \bibinfo {author} {\bibfnamefont
  {P.}~\bibnamefont {Jarillo-Herrero}},\ }\href {\doibase
  10.1038/s41565-021-01059-z} {\bibfield  {journal} {\bibinfo  {journal}
  {Nature Nanotechnology}\ }\textbf {\bibinfo {volume} {17}},\ \bibinfo {pages}
  {367} (\bibinfo {year} {2022})}\BibitemShut {NoStop}%
\bibitem [{\citenamefont {Meng}\ \emph {et~al.}(2022)\citenamefont {Meng},
  \citenamefont {Wu}, \citenamefont {Bian}, \citenamefont {Pan}, \citenamefont
  {Dong}, \citenamefont {Zhao}, \citenamefont {Chen}, \citenamefont {Wu},
  \citenamefont {Sun}, \citenamefont {Fu}, \citenamefont {Liu}, \citenamefont
  {Shi}, \citenamefont {Zhang}, \citenamefont {Zhang}, \citenamefont {Liu},\
  and\ \citenamefont {Liu}}]{MoS2.NC}%
  \BibitemOpen
  \bibfield  {author} {\bibinfo {author} {\bibfnamefont {P.}~\bibnamefont
  {Meng}}, \bibinfo {author} {\bibfnamefont {Y.}~\bibnamefont {Wu}}, \bibinfo
  {author} {\bibfnamefont {R.}~\bibnamefont {Bian}}, \bibinfo {author}
  {\bibfnamefont {E.}~\bibnamefont {Pan}}, \bibinfo {author} {\bibfnamefont
  {B.}~\bibnamefont {Dong}}, \bibinfo {author} {\bibfnamefont {X.}~\bibnamefont
  {Zhao}}, \bibinfo {author} {\bibfnamefont {J.}~\bibnamefont {Chen}}, \bibinfo
  {author} {\bibfnamefont {L.}~\bibnamefont {Wu}}, \bibinfo {author}
  {\bibfnamefont {Y.}~\bibnamefont {Sun}}, \bibinfo {author} {\bibfnamefont
  {Q.}~\bibnamefont {Fu}}, \bibinfo {author} {\bibfnamefont {Q.}~\bibnamefont
  {Liu}}, \bibinfo {author} {\bibfnamefont {D.}~\bibnamefont {Shi}}, \bibinfo
  {author} {\bibfnamefont {Q.}~\bibnamefont {Zhang}}, \bibinfo {author}
  {\bibfnamefont {Y.-W.}\ \bibnamefont {Zhang}}, \bibinfo {author}
  {\bibfnamefont {Z.}~\bibnamefont {Liu}}, \ and\ \bibinfo {author}
  {\bibfnamefont {F.}~\bibnamefont {Liu}},\ }\href {\doibase
  10.1038/s41467-022-35339-6} {\bibfield  {journal} {\bibinfo  {journal}
  {Nature Communications}\ }\textbf {\bibinfo {volume} {13}},\ \bibinfo {pages}
  {7696} (\bibinfo {year} {2022})}\BibitemShut {NoStop}%
\bibitem [{\citenamefont {Deb}\ \emph {et~al.}(2022)\citenamefont {Deb},
  \citenamefont {Cao}, \citenamefont {Raab}, \citenamefont {Watanabe},
  \citenamefont {Taniguchi}, \citenamefont {Goldstein}, \citenamefont {Kronik},
  \citenamefont {Urbakh}, \citenamefont {Hod},\ and\ \citenamefont
  {Ben~Shalom}}]{MoS2.Nature2022}%
  \BibitemOpen
  \bibfield  {author} {\bibinfo {author} {\bibfnamefont {S.}~\bibnamefont
  {Deb}}, \bibinfo {author} {\bibfnamefont {W.}~\bibnamefont {Cao}}, \bibinfo
  {author} {\bibfnamefont {N.}~\bibnamefont {Raab}}, \bibinfo {author}
  {\bibfnamefont {K.}~\bibnamefont {Watanabe}}, \bibinfo {author}
  {\bibfnamefont {T.}~\bibnamefont {Taniguchi}}, \bibinfo {author}
  {\bibfnamefont {M.}~\bibnamefont {Goldstein}}, \bibinfo {author}
  {\bibfnamefont {L.}~\bibnamefont {Kronik}}, \bibinfo {author} {\bibfnamefont
  {M.}~\bibnamefont {Urbakh}}, \bibinfo {author} {\bibfnamefont
  {O.}~\bibnamefont {Hod}}, \ and\ \bibinfo {author} {\bibfnamefont
  {M.}~\bibnamefont {Ben~Shalom}},\ }\href {\doibase
  10.1038/s41586-022-05341-5} {\bibfield  {journal} {\bibinfo  {journal}
  {Nature}\ }\textbf {\bibinfo {volume} {612}},\ \bibinfo {pages} {465}
  (\bibinfo {year} {2022})}\BibitemShut {NoStop}%
\bibitem [{\citenamefont {Fei}\ \emph {et~al.}(2018)\citenamefont {Fei},
  \citenamefont {Zhao}, \citenamefont {Palomaki}, \citenamefont {Sun},
  \citenamefont {Miller}, \citenamefont {Zhao}, \citenamefont {Yan},
  \citenamefont {Xu},\ and\ \citenamefont {Cobden}}]{WTe2.Nature}%
  \BibitemOpen
  \bibfield  {author} {\bibinfo {author} {\bibfnamefont {Z.}~\bibnamefont
  {Fei}}, \bibinfo {author} {\bibfnamefont {W.}~\bibnamefont {Zhao}}, \bibinfo
  {author} {\bibfnamefont {T.~A.}\ \bibnamefont {Palomaki}}, \bibinfo {author}
  {\bibfnamefont {B.}~\bibnamefont {Sun}}, \bibinfo {author} {\bibfnamefont
  {M.~K.}\ \bibnamefont {Miller}}, \bibinfo {author} {\bibfnamefont
  {Z.}~\bibnamefont {Zhao}}, \bibinfo {author} {\bibfnamefont {J.}~\bibnamefont
  {Yan}}, \bibinfo {author} {\bibfnamefont {X.}~\bibnamefont {Xu}}, \ and\
  \bibinfo {author} {\bibfnamefont {D.~H.}\ \bibnamefont {Cobden}},\ }\href
  {\doibase 10.1038/s41586-018-0336-3} {\bibfield  {journal} {\bibinfo
  {journal} {Nature}\ }\textbf {\bibinfo {volume} {560}},\ \bibinfo {pages}
  {336} (\bibinfo {year} {2018})}\BibitemShut {NoStop}%
\bibitem [{\citenamefont {Qing}\ \emph {et~al.}(2018)\citenamefont {Qing},
  \citenamefont {Wu},\ and\ \citenamefont {Ju}}]{WTe2.JPCL}%
  \BibitemOpen
  \bibfield  {author} {\bibinfo {author} {\bibfnamefont {Y.}~\bibnamefont
  {Qing}}, \bibinfo {author} {\bibfnamefont {M.}~\bibnamefont {Wu}}, \ and\
  \bibinfo {author} {\bibfnamefont {L.}~\bibnamefont {Ju}},\ }\href {\doibase
  10.1021/acs.jpclett.8b03654} {\bibfield  {journal} {\bibinfo  {journal} {J.
  Phys. Chem. Lett.}\ }\textbf {\bibinfo {volume} {9}},\ \bibinfo {pages}
  {7160–7164} (\bibinfo {year} {2018})}\BibitemShut {NoStop}%
\bibitem [{\citenamefont {Xiao}\ \emph {et~al.}(2020)\citenamefont {Xiao},
  \citenamefont {Wang}, \citenamefont {Wang}, \citenamefont {Pemmaraju},
  \citenamefont {Wang}, \citenamefont {Muscher}, \citenamefont {Sie},
  \citenamefont {Nyby}, \citenamefont {Devereaux}, \citenamefont {Qian},
  \citenamefont {Zhang},\ and\ \citenamefont {Lindenberg}}]{WTe2.NP}%
  \BibitemOpen
  \bibfield  {author} {\bibinfo {author} {\bibfnamefont {J.}~\bibnamefont
  {Xiao}}, \bibinfo {author} {\bibfnamefont {Y.}~\bibnamefont {Wang}}, \bibinfo
  {author} {\bibfnamefont {H.}~\bibnamefont {Wang}}, \bibinfo {author}
  {\bibfnamefont {C.~D.}\ \bibnamefont {Pemmaraju}}, \bibinfo {author}
  {\bibfnamefont {S.}~\bibnamefont {Wang}}, \bibinfo {author} {\bibfnamefont
  {P.}~\bibnamefont {Muscher}}, \bibinfo {author} {\bibfnamefont {E.~J.}\
  \bibnamefont {Sie}}, \bibinfo {author} {\bibfnamefont {C.~M.}\ \bibnamefont
  {Nyby}}, \bibinfo {author} {\bibfnamefont {T.~P.}\ \bibnamefont {Devereaux}},
  \bibinfo {author} {\bibfnamefont {X.}~\bibnamefont {Qian}}, \bibinfo {author}
  {\bibfnamefont {X.}~\bibnamefont {Zhang}}, \ and\ \bibinfo {author}
  {\bibfnamefont {A.~M.}\ \bibnamefont {Lindenberg}},\ }\href {\doibase
  10.1038/s41567-020-0947-0} {\bibfield  {journal} {\bibinfo  {journal} {Nature
  Physics}\ }\textbf {\bibinfo {volume} {16}},\ \bibinfo {pages} {1028}
  (\bibinfo {year} {2020})}\BibitemShut {NoStop}%
\bibitem [{\citenamefont {Wan}\ \emph {et~al.}(2022)\citenamefont {Wan},
  \citenamefont {Hu}, \citenamefont {Mao}, \citenamefont {Fu}, \citenamefont
  {Yuan}, \citenamefont {Song}, \citenamefont {Gan}, \citenamefont {Xu},
  \citenamefont {Xue}, \citenamefont {Cheng}, \citenamefont {Huang},
  \citenamefont {Yang}, \citenamefont {Dai}, \citenamefont {Zeng},\ and\
  \citenamefont {Kan}}]{ReS2.PRL}%
  \BibitemOpen
  \bibfield  {author} {\bibinfo {author} {\bibfnamefont {Y.}~\bibnamefont
  {Wan}}, \bibinfo {author} {\bibfnamefont {T.}~\bibnamefont {Hu}}, \bibinfo
  {author} {\bibfnamefont {X.}~\bibnamefont {Mao}}, \bibinfo {author}
  {\bibfnamefont {J.}~\bibnamefont {Fu}}, \bibinfo {author} {\bibfnamefont
  {K.}~\bibnamefont {Yuan}}, \bibinfo {author} {\bibfnamefont {Y.}~\bibnamefont
  {Song}}, \bibinfo {author} {\bibfnamefont {X.}~\bibnamefont {Gan}}, \bibinfo
  {author} {\bibfnamefont {X.}~\bibnamefont {Xu}}, \bibinfo {author}
  {\bibfnamefont {M.}~\bibnamefont {Xue}}, \bibinfo {author} {\bibfnamefont
  {X.}~\bibnamefont {Cheng}}, \bibinfo {author} {\bibfnamefont
  {C.}~\bibnamefont {Huang}}, \bibinfo {author} {\bibfnamefont
  {J.}~\bibnamefont {Yang}}, \bibinfo {author} {\bibfnamefont {L.}~\bibnamefont
  {Dai}}, \bibinfo {author} {\bibfnamefont {H.}~\bibnamefont {Zeng}}, \ and\
  \bibinfo {author} {\bibfnamefont {E.}~\bibnamefont {Kan}},\ }\href {\doibase
  10.1103/PhysRevLett.128.067601} {\bibfield  {journal} {\bibinfo  {journal}
  {Phys. Rev. Lett.}\ }\textbf {\bibinfo {volume} {128}},\ \bibinfo {pages}
  {067601} (\bibinfo {year} {2022})}\BibitemShut {NoStop}%
\bibitem [{\citenamefont {Jindal}\ \emph {et~al.}(2023)\citenamefont {Jindal},
  \citenamefont {Saha}, \citenamefont {Li}, \citenamefont {Taniguchi},
  \citenamefont {Watanabe}, \citenamefont {Hone}, \citenamefont {Birol},
  \citenamefont {Fernandes}, \citenamefont {Dean}, \citenamefont {Pasupathy},\
  and\ \citenamefont {Rhodes}}]{T1-MoTe2.Nature}%
  \BibitemOpen
  \bibfield  {author} {\bibinfo {author} {\bibfnamefont {A.}~\bibnamefont
  {Jindal}}, \bibinfo {author} {\bibfnamefont {A.}~\bibnamefont {Saha}},
  \bibinfo {author} {\bibfnamefont {Z.}~\bibnamefont {Li}}, \bibinfo {author}
  {\bibfnamefont {T.}~\bibnamefont {Taniguchi}}, \bibinfo {author}
  {\bibfnamefont {K.}~\bibnamefont {Watanabe}}, \bibinfo {author}
  {\bibfnamefont {J.~C.}\ \bibnamefont {Hone}}, \bibinfo {author}
  {\bibfnamefont {T.}~\bibnamefont {Birol}}, \bibinfo {author} {\bibfnamefont
  {R.~M.}\ \bibnamefont {Fernandes}}, \bibinfo {author} {\bibfnamefont {C.~R.}\
  \bibnamefont {Dean}}, \bibinfo {author} {\bibfnamefont {A.~N.}\ \bibnamefont
  {Pasupathy}}, \ and\ \bibinfo {author} {\bibfnamefont {D.~A.}\ \bibnamefont
  {Rhodes}},\ }\href {\doibase 10.1038/s41586-022-05521-3} {\bibfield
  {journal} {\bibinfo  {journal} {Nature}\ }\textbf {\bibinfo {volume} {613}},\
  \bibinfo {pages} {48} (\bibinfo {year} {2023})}\BibitemShut {NoStop}%
\bibitem [{\citenamefont {Ji}\ \emph {et~al.}(2023)\citenamefont {Ji},
  \citenamefont {Yu}, \citenamefont {Xu},\ and\ \citenamefont
  {Xiang}}]{Stacking.arXiv}%
  \BibitemOpen
  \bibfield  {author} {\bibinfo {author} {\bibfnamefont {J.}~\bibnamefont
  {Ji}}, \bibinfo {author} {\bibfnamefont {G.}~\bibnamefont {Yu}}, \bibinfo
  {author} {\bibfnamefont {C.}~\bibnamefont {Xu}}, \ and\ \bibinfo {author}
  {\bibfnamefont {H.~J.}\ \bibnamefont {Xiang}},\ }\href {\doibase
  10.1103/PhysRevLett.130.146801} {\bibfield  {journal} {\bibinfo  {journal}
  {Phys. Rev. Lett.}\ }\textbf {\bibinfo {volume} {130}},\ \bibinfo {pages}
  {146801} (\bibinfo {year} {2023})}\BibitemShut {NoStop}%
\bibitem [{\citenamefont {Anderson}\ and\ \citenamefont
  {Blount}(1965)}]{Anderson1965}%
  \BibitemOpen
  \bibfield  {author} {\bibinfo {author} {\bibfnamefont {P.~W.}\ \bibnamefont
  {Anderson}}\ and\ \bibinfo {author} {\bibfnamefont {E.~I.}\ \bibnamefont
  {Blount}},\ }\href {\doibase 10.1103/PhysRevLett.14.217} {\bibfield
  {journal} {\bibinfo  {journal} {Phys. Rev. Lett.}\ }\textbf {\bibinfo
  {volume} {14}},\ \bibinfo {pages} {217} (\bibinfo {year} {1965})}\BibitemShut
  {NoStop}%
\bibitem [{\citenamefont {Shi}\ \emph {et~al.}(2013)\citenamefont {Shi},
  \citenamefont {Guo}, \citenamefont {Wang}, \citenamefont {Princep},
  \citenamefont {Khalyavin}, \citenamefont {Manuel}, \citenamefont {Michiue},
  \citenamefont {Sato}, \citenamefont {Tsuda}, \citenamefont {Yu},
  \citenamefont {Arai}, \citenamefont {Shirako}, \citenamefont {Akaogi},
  \citenamefont {Wang}, \citenamefont {Yamaura},\ and\ \citenamefont
  {Boothroyd}}]{LiOsO.NM}%
  \BibitemOpen
  \bibfield  {author} {\bibinfo {author} {\bibfnamefont {Y.}~\bibnamefont
  {Shi}}, \bibinfo {author} {\bibfnamefont {Y.}~\bibnamefont {Guo}}, \bibinfo
  {author} {\bibfnamefont {X.}~\bibnamefont {Wang}}, \bibinfo {author}
  {\bibfnamefont {A.~J.}\ \bibnamefont {Princep}}, \bibinfo {author}
  {\bibfnamefont {D.}~\bibnamefont {Khalyavin}}, \bibinfo {author}
  {\bibfnamefont {P.}~\bibnamefont {Manuel}}, \bibinfo {author} {\bibfnamefont
  {Y.}~\bibnamefont {Michiue}}, \bibinfo {author} {\bibfnamefont
  {A.}~\bibnamefont {Sato}}, \bibinfo {author} {\bibfnamefont {K.}~\bibnamefont
  {Tsuda}}, \bibinfo {author} {\bibfnamefont {S.}~\bibnamefont {Yu}}, \bibinfo
  {author} {\bibfnamefont {M.}~\bibnamefont {Arai}}, \bibinfo {author}
  {\bibfnamefont {Y.}~\bibnamefont {Shirako}}, \bibinfo {author} {\bibfnamefont
  {M.}~\bibnamefont {Akaogi}}, \bibinfo {author} {\bibfnamefont
  {N.}~\bibnamefont {Wang}}, \bibinfo {author} {\bibfnamefont {K.}~\bibnamefont
  {Yamaura}}, \ and\ \bibinfo {author} {\bibfnamefont {A.~T.}\ \bibnamefont
  {Boothroyd}},\ }\href {\doibase 10.1038/nmat3754} {\bibfield  {journal}
  {\bibinfo  {journal} {Nature Materials}\ }\textbf {\bibinfo {volume} {12}},\
  \bibinfo {pages} {1024} (\bibinfo {year} {2013})}\BibitemShut {NoStop}%
\bibitem [{\citenamefont {Xiang}(2014)}]{LiOsO3.PhysRevB.90.094108}%
  \BibitemOpen
  \bibfield  {author} {\bibinfo {author} {\bibfnamefont {H.~J.}\ \bibnamefont
  {Xiang}},\ }\href {\doibase 10.1103/PhysRevB.90.094108} {\bibfield  {journal}
  {\bibinfo  {journal} {Phys. Rev. B}\ }\textbf {\bibinfo {volume} {90}},\
  \bibinfo {pages} {094108} (\bibinfo {year} {2014})}\BibitemShut {NoStop}%
\bibitem [{\citenamefont {Rischau}\ \emph {et~al.}(2017)\citenamefont
  {Rischau}, \citenamefont {Lin}, \citenamefont {Grams}, \citenamefont {Finck},
  \citenamefont {Harms}, \citenamefont {Engelmayer}, \citenamefont {Lorenz},
  \citenamefont {Gallais}, \citenamefont {Fauqué}, \citenamefont {Hemberger},\
  and\ \citenamefont {Behnia}}]{SrTiO.supe.NP}%
  \BibitemOpen
  \bibfield  {author} {\bibinfo {author} {\bibfnamefont {C.~W.}\ \bibnamefont
  {Rischau}}, \bibinfo {author} {\bibfnamefont {X.}~\bibnamefont {Lin}},
  \bibinfo {author} {\bibfnamefont {C.~P.}\ \bibnamefont {Grams}}, \bibinfo
  {author} {\bibfnamefont {D.}~\bibnamefont {Finck}}, \bibinfo {author}
  {\bibfnamefont {S.}~\bibnamefont {Harms}}, \bibinfo {author} {\bibfnamefont
  {J.}~\bibnamefont {Engelmayer}}, \bibinfo {author} {\bibfnamefont
  {T.}~\bibnamefont {Lorenz}}, \bibinfo {author} {\bibfnamefont
  {Y.}~\bibnamefont {Gallais}}, \bibinfo {author} {\bibfnamefont
  {B.}~\bibnamefont {Fauqué}}, \bibinfo {author} {\bibfnamefont
  {J.}~\bibnamefont {Hemberger}}, \ and\ \bibinfo {author} {\bibfnamefont
  {K.}~\bibnamefont {Behnia}},\ }\href {\doibase 10.1038/nphys4085} {\bibfield
  {journal} {\bibinfo  {journal} {Nature Physics}\ }\textbf {\bibinfo {volume}
  {13}},\ \bibinfo {pages} {643} (\bibinfo {year} {2017})}\BibitemShut
  {NoStop}%
\bibitem [{\citenamefont {Enderlein}\ \emph {et~al.}(2020)\citenamefont
  {Enderlein}, \citenamefont {de~Oliveira}, \citenamefont {Tompsett},
  \citenamefont {Saitovitch}, \citenamefont {Saxena}, \citenamefont
  {Lonzarich},\ and\ \citenamefont {Rowley}}]{SrTiO.supe.NC}%
  \BibitemOpen
  \bibfield  {author} {\bibinfo {author} {\bibfnamefont {C.}~\bibnamefont
  {Enderlein}}, \bibinfo {author} {\bibfnamefont {J.~F.}\ \bibnamefont
  {de~Oliveira}}, \bibinfo {author} {\bibfnamefont {D.~A.}\ \bibnamefont
  {Tompsett}}, \bibinfo {author} {\bibfnamefont {E.~B.}\ \bibnamefont
  {Saitovitch}}, \bibinfo {author} {\bibfnamefont {S.~S.}\ \bibnamefont
  {Saxena}}, \bibinfo {author} {\bibfnamefont {G.~G.}\ \bibnamefont
  {Lonzarich}}, \ and\ \bibinfo {author} {\bibfnamefont {S.~E.}\ \bibnamefont
  {Rowley}},\ }\href {\doibase 10.1038/s41467-020-18438-0} {\bibfield
  {journal} {\bibinfo  {journal} {Nature Communications}\ }\textbf {\bibinfo
  {volume} {11}},\ \bibinfo {pages} {4852} (\bibinfo {year}
  {2020})}\BibitemShut {NoStop}%
\bibitem [{\citenamefont {Cai}\ \emph {et~al.}(2021)\citenamefont {Cai},
  \citenamefont {He}, \citenamefont {Li}, \citenamefont {Zhang}, \citenamefont
  {Zhang}, \citenamefont {Chung}, \citenamefont {Bhowmick}, \citenamefont
  {Wolverton}, \citenamefont {Kanatzidis},\ and\ \citenamefont
  {Deemyad}}]{HgTeX.NC}%
  \BibitemOpen
  \bibfield  {author} {\bibinfo {author} {\bibfnamefont {W.}~\bibnamefont
  {Cai}}, \bibinfo {author} {\bibfnamefont {J.}~\bibnamefont {He}}, \bibinfo
  {author} {\bibfnamefont {H.}~\bibnamefont {Li}}, \bibinfo {author}
  {\bibfnamefont {R.}~\bibnamefont {Zhang}}, \bibinfo {author} {\bibfnamefont
  {D.}~\bibnamefont {Zhang}}, \bibinfo {author} {\bibfnamefont {D.~Y.}\
  \bibnamefont {Chung}}, \bibinfo {author} {\bibfnamefont {T.}~\bibnamefont
  {Bhowmick}}, \bibinfo {author} {\bibfnamefont {C.}~\bibnamefont {Wolverton}},
  \bibinfo {author} {\bibfnamefont {M.~G.}\ \bibnamefont {Kanatzidis}}, \ and\
  \bibinfo {author} {\bibfnamefont {S.}~\bibnamefont {Deemyad}},\ }\href
  {\doibase 10.1038/s41467-021-21836-7} {\bibfield  {journal} {\bibinfo
  {journal} {Nature Communications}\ }\textbf {\bibinfo {volume} {12}},\
  \bibinfo {pages} {1509} (\bibinfo {year} {2021})}\BibitemShut {NoStop}%
\bibitem [{\citenamefont {Wang}\ \emph {et~al.}(2019)\citenamefont {Wang},
  \citenamefont {Wieder}, \citenamefont {Li}, \citenamefont {Yan},\ and\
  \citenamefont {Bernevig}}]{MoTe2.wzj.PRL}%
  \BibitemOpen
  \bibfield  {author} {\bibinfo {author} {\bibfnamefont {Z.}~\bibnamefont
  {Wang}}, \bibinfo {author} {\bibfnamefont {B.~J.}\ \bibnamefont {Wieder}},
  \bibinfo {author} {\bibfnamefont {J.}~\bibnamefont {Li}}, \bibinfo {author}
  {\bibfnamefont {B.}~\bibnamefont {Yan}}, \ and\ \bibinfo {author}
  {\bibfnamefont {B.~A.}\ \bibnamefont {Bernevig}},\ }\href {\doibase
  10.1103/PhysRevLett.123.186401} {\bibfield  {journal} {\bibinfo  {journal}
  {Phys. Rev. Lett.}\ }\textbf {\bibinfo {volume} {123}},\ \bibinfo {pages}
  {186401} (\bibinfo {year} {2019})}\BibitemShut {NoStop}%
\bibitem [{\citenamefont {Huang}\ \emph {et~al.}(2019)\citenamefont {Huang},
  \citenamefont {Joon~Lim}, \citenamefont {Singh}, \citenamefont {Kim},
  \citenamefont {Zhang}, \citenamefont {Kim}, \citenamefont {Chu},
  \citenamefont {Rabe}, \citenamefont {Vanderbilt},\ and\ \citenamefont
  {Cheong}}]{MoTe2-T0.NC}%
  \BibitemOpen
  \bibfield  {author} {\bibinfo {author} {\bibfnamefont {F.-T.}\ \bibnamefont
  {Huang}}, \bibinfo {author} {\bibfnamefont {S.}~\bibnamefont {Joon~Lim}},
  \bibinfo {author} {\bibfnamefont {S.}~\bibnamefont {Singh}}, \bibinfo
  {author} {\bibfnamefont {J.}~\bibnamefont {Kim}}, \bibinfo {author}
  {\bibfnamefont {L.}~\bibnamefont {Zhang}}, \bibinfo {author} {\bibfnamefont
  {J.-W.}\ \bibnamefont {Kim}}, \bibinfo {author} {\bibfnamefont {M.-W.}\
  \bibnamefont {Chu}}, \bibinfo {author} {\bibfnamefont {K.~M.}\ \bibnamefont
  {Rabe}}, \bibinfo {author} {\bibfnamefont {D.}~\bibnamefont {Vanderbilt}}, \
  and\ \bibinfo {author} {\bibfnamefont {S.-W.}\ \bibnamefont {Cheong}},\
  }\href {\doibase 10.1038/s41467-019-11949-5} {\bibfield  {journal} {\bibinfo
  {journal} {Nature Communications}\ }\textbf {\bibinfo {volume} {10}},\
  \bibinfo {pages} {4211} (\bibinfo {year} {2019})}\BibitemShut {NoStop}%
\bibitem [{\citenamefont {Luo}\ \emph {et~al.}(2017)\citenamefont {Luo},
  \citenamefont {Xu},\ and\ \citenamefont {Xiang}}]{CrN.PRB}%
  \BibitemOpen
  \bibfield  {author} {\bibinfo {author} {\bibfnamefont {W.}~\bibnamefont
  {Luo}}, \bibinfo {author} {\bibfnamefont {K.}~\bibnamefont {Xu}}, \ and\
  \bibinfo {author} {\bibfnamefont {H.}~\bibnamefont {Xiang}},\ }\href
  {\doibase 10.1103/PhysRevB.96.235415} {\bibfield  {journal} {\bibinfo
  {journal} {Phys. Rev. B}\ }\textbf {\bibinfo {volume} {96}},\ \bibinfo
  {pages} {235415} (\bibinfo {year} {2017})}\BibitemShut {NoStop}%
\bibitem [{\citenamefont {Lu}\ \emph {et~al.}(2019)\citenamefont {Lu},
  \citenamefont {Chen}, \citenamefont {Luo}, \citenamefont {\'I\~niguez},
  \citenamefont {Bellaiche},\ and\ \citenamefont {Xiang}}]{LiOsO3film.PRL}%
  \BibitemOpen
  \bibfield  {author} {\bibinfo {author} {\bibfnamefont {J.}~\bibnamefont
  {Lu}}, \bibinfo {author} {\bibfnamefont {G.}~\bibnamefont {Chen}}, \bibinfo
  {author} {\bibfnamefont {W.}~\bibnamefont {Luo}}, \bibinfo {author}
  {\bibfnamefont {J.}~\bibnamefont {\'I\~niguez}}, \bibinfo {author}
  {\bibfnamefont {L.}~\bibnamefont {Bellaiche}}, \ and\ \bibinfo {author}
  {\bibfnamefont {H.}~\bibnamefont {Xiang}},\ }\href {\doibase
  10.1103/PhysRevLett.122.227601} {\bibfield  {journal} {\bibinfo  {journal}
  {Phys. Rev. Lett.}\ }\textbf {\bibinfo {volume} {122}},\ \bibinfo {pages}
  {227601} (\bibinfo {year} {2019})}\BibitemShut {NoStop}%
\bibitem [{\citenamefont {Ma}\ \emph {et~al.}(2021)\citenamefont {Ma},
  \citenamefont {Lyu}, \citenamefont {Hao}, \citenamefont {Zhao}, \citenamefont
  {Qian}, \citenamefont {Yan},\ and\ \citenamefont {Su}}]{M1M1P2X6.SB}%
  \BibitemOpen
  \bibfield  {author} {\bibinfo {author} {\bibfnamefont {X.-Y.}\ \bibnamefont
  {Ma}}, \bibinfo {author} {\bibfnamefont {H.-Y.}\ \bibnamefont {Lyu}},
  \bibinfo {author} {\bibfnamefont {K.-R.}\ \bibnamefont {Hao}}, \bibinfo
  {author} {\bibfnamefont {Y.-M.}\ \bibnamefont {Zhao}}, \bibinfo {author}
  {\bibfnamefont {X.}~\bibnamefont {Qian}}, \bibinfo {author} {\bibfnamefont
  {Q.-B.}\ \bibnamefont {Yan}}, \ and\ \bibinfo {author} {\bibfnamefont
  {G.}~\bibnamefont {Su}},\ }\href {\doibase
  https://doi.org/10.1016/j.scib.2020.09.010} {\bibfield  {journal} {\bibinfo
  {journal} {Science Bulletin}\ }\textbf {\bibinfo {volume} {66}},\ \bibinfo
  {pages} {233} (\bibinfo {year} {2021})}\BibitemShut {NoStop}%
\bibitem [{\citenamefont {de~la Barrera}\ \emph {et~al.}(2021)\citenamefont
  {de~la Barrera}, \citenamefont {Cao}, \citenamefont {Gao}, \citenamefont
  {Gao}, \citenamefont {Bheemarasetty}, \citenamefont {Yan}, \citenamefont
  {Mandrus}, \citenamefont {Zhu}, \citenamefont {Xiao},\ and\ \citenamefont
  {Hunt}}]{WTe2.doping.NC}%
  \BibitemOpen
  \bibfield  {author} {\bibinfo {author} {\bibfnamefont {S.~C.}\ \bibnamefont
  {de~la Barrera}}, \bibinfo {author} {\bibfnamefont {Q.}~\bibnamefont {Cao}},
  \bibinfo {author} {\bibfnamefont {Y.}~\bibnamefont {Gao}}, \bibinfo {author}
  {\bibfnamefont {Y.}~\bibnamefont {Gao}}, \bibinfo {author} {\bibfnamefont
  {V.~S.}\ \bibnamefont {Bheemarasetty}}, \bibinfo {author} {\bibfnamefont
  {J.}~\bibnamefont {Yan}}, \bibinfo {author} {\bibfnamefont {D.~G.}\
  \bibnamefont {Mandrus}}, \bibinfo {author} {\bibfnamefont {W.}~\bibnamefont
  {Zhu}}, \bibinfo {author} {\bibfnamefont {D.}~\bibnamefont {Xiao}}, \ and\
  \bibinfo {author} {\bibfnamefont {B.~M.}\ \bibnamefont {Hunt}},\ }\href
  {\doibase 10.1038/s41467-021-25587-3} {\bibfield  {journal} {\bibinfo
  {journal} {Nature Communications}\ }\textbf {\bibinfo {volume} {12}},\
  \bibinfo {pages} {5298} (\bibinfo {year} {2021})}\BibitemShut {NoStop}%
\bibitem [{\citenamefont {Cohen}(1992)}]{BaTiO.Nature}%
  \BibitemOpen
  \bibfield  {author} {\bibinfo {author} {\bibfnamefont {R.~E.}\ \bibnamefont
  {Cohen}},\ }\href {\doibase 10.1038/358136a0} {\bibfield  {journal} {\bibinfo
   {journal} {Nature}\ }\textbf {\bibinfo {volume} {358}},\ \bibinfo {pages}
  {136} (\bibinfo {year} {1992})}\BibitemShut {NoStop}%
\bibitem [{\citenamefont {Ke}\ \emph {et~al.}(2021)\citenamefont {Ke},
  \citenamefont {Huang},\ and\ \citenamefont {Liu}}]{In2Se3.MH}%
  \BibitemOpen
  \bibfield  {author} {\bibinfo {author} {\bibfnamefont {C.}~\bibnamefont
  {Ke}}, \bibinfo {author} {\bibfnamefont {J.}~\bibnamefont {Huang}}, \ and\
  \bibinfo {author} {\bibfnamefont {S.}~\bibnamefont {Liu}},\ }\href {\doibase
  10.1039/D1MH01556G} {\bibfield  {journal} {\bibinfo  {journal} {Mater.
  Horiz.}\ }\textbf {\bibinfo {volume} {8}},\ \bibinfo {pages} {3387} (\bibinfo
  {year} {2021})}\BibitemShut {NoStop}%
\bibitem [{\citenamefont {Saxena}(2004)}]{SC.Nature2004}%
  \BibitemOpen
  \bibfield  {author} {\bibinfo {author} {\bibfnamefont {P.}~\bibnamefont
  {Saxena}, \bibfnamefont {S.and~Monthoux}},\ }\href {\doibase 10.1038/427799a}
  {\bibfield  {journal} {\bibinfo  {journal} {Nature}\ }\textbf {\bibinfo
  {volume} {427}},\ \bibinfo {pages} {799} (\bibinfo {year}
  {2004})}\BibitemShut {NoStop}%
\bibitem [{\citenamefont {Mineev}\ and\ \citenamefont
  {Yoshioka}(2010)}]{OP.PRB2010}%
  \BibitemOpen
  \bibfield  {author} {\bibinfo {author} {\bibfnamefont {V.~P.}\ \bibnamefont
  {Mineev}}\ and\ \bibinfo {author} {\bibfnamefont {Y.}~\bibnamefont
  {Yoshioka}},\ }\href {\doibase 10.1103/PhysRevB.81.094525} {\bibfield
  {journal} {\bibinfo  {journal} {Phys. Rev. B}\ }\textbf {\bibinfo {volume}
  {81}},\ \bibinfo {pages} {094525} (\bibinfo {year} {2010})}\BibitemShut
  {NoStop}%
\bibitem [{\citenamefont {Edelstein}(2011)}]{OP.PRB2011}%
  \BibitemOpen
  \bibfield  {author} {\bibinfo {author} {\bibfnamefont {V.~M.}\ \bibnamefont
  {Edelstein}},\ }\href {\doibase 10.1103/PhysRevB.83.113109} {\bibfield
  {journal} {\bibinfo  {journal} {Phys. Rev. B}\ }\textbf {\bibinfo {volume}
  {83}},\ \bibinfo {pages} {113109} (\bibinfo {year} {2011})}\BibitemShut
  {NoStop}%
\bibitem [{\citenamefont {Edelstein}(1995)}]{ME.PRL1995}%
  \BibitemOpen
  \bibfield  {author} {\bibinfo {author} {\bibfnamefont {V.~M.}\ \bibnamefont
  {Edelstein}},\ }\href {\doibase 10.1103/PhysRevLett.75.2004} {\bibfield
  {journal} {\bibinfo  {journal} {Phys. Rev. Lett.}\ }\textbf {\bibinfo
  {volume} {75}},\ \bibinfo {pages} {2004} (\bibinfo {year}
  {1995})}\BibitemShut {NoStop}%
\bibitem [{\citenamefont {Kanasugi}\ and\ \citenamefont
  {Yanase}(2018)}]{ME.PRB2018}%
  \BibitemOpen
  \bibfield  {author} {\bibinfo {author} {\bibfnamefont {S.}~\bibnamefont
  {Kanasugi}}\ and\ \bibinfo {author} {\bibfnamefont {Y.}~\bibnamefont
  {Yanase}},\ }\href {\doibase 10.1103/PhysRevB.98.024521} {\bibfield
  {journal} {\bibinfo  {journal} {Phys. Rev. B}\ }\textbf {\bibinfo {volume}
  {98}},\ \bibinfo {pages} {024521} (\bibinfo {year} {2018})}\BibitemShut
  {NoStop}%
\bibitem [{\citenamefont {Bl\"ochl}(1994)}]{paw1}%
  \BibitemOpen
  \bibfield  {author} {\bibinfo {author} {\bibfnamefont {P.~E.}\ \bibnamefont
  {Bl\"ochl}},\ }\href {\doibase 10.1103/PhysRevB.50.17953} {\bibfield
  {journal} {\bibinfo  {journal} {Phys. Rev. B}\ }\textbf {\bibinfo {volume}
  {50}},\ \bibinfo {pages} {17953} (\bibinfo {year} {1994})}\BibitemShut
  {NoStop}%
\bibitem [{\citenamefont {Kresse}\ and\ \citenamefont {Joubert}(1999)}]{paw2}%
  \BibitemOpen
  \bibfield  {author} {\bibinfo {author} {\bibfnamefont {G.}~\bibnamefont
  {Kresse}}\ and\ \bibinfo {author} {\bibfnamefont {D.}~\bibnamefont
  {Joubert}},\ }\href {\doibase 10.1103/PhysRevB.59.1758} {\bibfield  {journal}
  {\bibinfo  {journal} {Phys. Rev. B}\ }\textbf {\bibinfo {volume} {59}},\
  \bibinfo {pages} {1758} (\bibinfo {year} {1999})}\BibitemShut {NoStop}%
\bibitem [{\citenamefont {Kresse}\ and\ \citenamefont
  {Furthmüller}(1996)}]{KRESSE199615}%
  \BibitemOpen
  \bibfield  {author} {\bibinfo {author} {\bibfnamefont {G.}~\bibnamefont
  {Kresse}}\ and\ \bibinfo {author} {\bibfnamefont {J.}~\bibnamefont
  {Furthmüller}},\ }\href {\doibase
  https://doi.org/10.1016/0927-0256(96)00008-0} {\bibfield  {journal} {\bibinfo
   {journal} {Computational Materials Science}\ }\textbf {\bibinfo {volume}
  {6}},\ \bibinfo {pages} {15 } (\bibinfo {year} {1996})}\BibitemShut {NoStop}%
\bibitem [{\citenamefont {Kresse}\ and\ \citenamefont
  {Furthm\"uller}(1996)}]{vasp}%
  \BibitemOpen
  \bibfield  {author} {\bibinfo {author} {\bibfnamefont {G.}~\bibnamefont
  {Kresse}}\ and\ \bibinfo {author} {\bibfnamefont {J.}~\bibnamefont
  {Furthm\"uller}},\ }\href {\doibase 10.1103/PhysRevB.54.11169} {\bibfield
  {journal} {\bibinfo  {journal} {Phys. Rev. B}\ }\textbf {\bibinfo {volume}
  {54}},\ \bibinfo {pages} {11169} (\bibinfo {year} {1996})}\BibitemShut
  {NoStop}%
\bibitem [{\citenamefont {Perdew}\ \emph {et~al.}(1996)\citenamefont {Perdew},
  \citenamefont {Burke},\ and\ \citenamefont {Ernzerhof}}]{pbe}%
  \BibitemOpen
  \bibfield  {author} {\bibinfo {author} {\bibfnamefont {J.~P.}\ \bibnamefont
  {Perdew}}, \bibinfo {author} {\bibfnamefont {K.}~\bibnamefont {Burke}}, \
  and\ \bibinfo {author} {\bibfnamefont {M.}~\bibnamefont {Ernzerhof}},\ }\href
  {\doibase 10.1103/PhysRevLett.77.3865} {\bibfield  {journal} {\bibinfo
  {journal} {Phys. Rev. Lett.}\ }\textbf {\bibinfo {volume} {77}},\ \bibinfo
  {pages} {3865} (\bibinfo {year} {1996})}\BibitemShut {NoStop}%
\bibitem [{\citenamefont {Grimme}\ \emph {et~al.}(2010)\citenamefont {Grimme},
  \citenamefont {Antony}, \citenamefont {Ehrlich},\ and\ \citenamefont
  {Krieg}}]{DFT-D3.JCP}%
  \BibitemOpen
  \bibfield  {author} {\bibinfo {author} {\bibfnamefont {S.}~\bibnamefont
  {Grimme}}, \bibinfo {author} {\bibfnamefont {J.}~\bibnamefont {Antony}},
  \bibinfo {author} {\bibfnamefont {S.}~\bibnamefont {Ehrlich}}, \ and\
  \bibinfo {author} {\bibfnamefont {H.}~\bibnamefont {Krieg}},\ }\href
  {\doibase 10.1063/1.3382344} {\bibfield  {journal} {\bibinfo  {journal} {The
  Journal of Chemical Physics}\ }\textbf {\bibinfo {volume} {132}},\ \bibinfo
  {pages} {154104} (\bibinfo {year} {2010})}\BibitemShut {NoStop}%
\bibitem [{\citenamefont {Grimme}\ \emph {et~al.}(2011)\citenamefont {Grimme},
  \citenamefont {Ehrlich},\ and\ \citenamefont {Goerigk}}]{DFT-D3.JCC}%
  \BibitemOpen
  \bibfield  {author} {\bibinfo {author} {\bibfnamefont {S.}~\bibnamefont
  {Grimme}}, \bibinfo {author} {\bibfnamefont {S.}~\bibnamefont {Ehrlich}}, \
  and\ \bibinfo {author} {\bibfnamefont {L.}~\bibnamefont {Goerigk}},\ }\href
  {\doibase https://doi.org/10.1002/jcc.21759} {\bibfield  {journal} {\bibinfo
  {journal} {Journal of Computational Chemistry}\ }\textbf {\bibinfo {volume}
  {32}},\ \bibinfo {pages} {1456} (\bibinfo {year} {2011})}\BibitemShut
  {NoStop}%
\bibitem [{\citenamefont {Neugebauer}\ and\ \citenamefont
  {Scheffler}(1992)}]{dipole}%
  \BibitemOpen
  \bibfield  {author} {\bibinfo {author} {\bibfnamefont {J.}~\bibnamefont
  {Neugebauer}}\ and\ \bibinfo {author} {\bibfnamefont {M.}~\bibnamefont
  {Scheffler}},\ }\href {\doibase 10.1103/PhysRevB.46.16067} {\bibfield
  {journal} {\bibinfo  {journal} {Phys. Rev. B}\ }\textbf {\bibinfo {volume}
  {46}},\ \bibinfo {pages} {16067} (\bibinfo {year} {1992})}\BibitemShut
  {NoStop}%
\bibitem [{\citenamefont {Heyd}\ \emph {et~al.}(2003)\citenamefont {Heyd},
  \citenamefont {Scuseria},\ and\ \citenamefont {Ernzerhof}}]{HSE.JCP}%
  \BibitemOpen
  \bibfield  {author} {\bibinfo {author} {\bibfnamefont {J.}~\bibnamefont
  {Heyd}}, \bibinfo {author} {\bibfnamefont {G.~E.}\ \bibnamefont {Scuseria}},
  \ and\ \bibinfo {author} {\bibfnamefont {M.}~\bibnamefont {Ernzerhof}},\
  }\href {\doibase 10.1063/1.1564060} {\bibfield  {journal} {\bibinfo
  {journal} {The Journal of Chemical Physics}\ }\textbf {\bibinfo {volume}
  {118}},\ \bibinfo {pages} {8207} (\bibinfo {year} {2003})}\BibitemShut
  {NoStop}%
\bibitem [{\citenamefont {Henkelman}\ \emph {et~al.}(2000)\citenamefont
  {Henkelman}, \citenamefont {Uberuaga},\ and\ \citenamefont
  {Jónsson}}]{CINEB}%
  \BibitemOpen
  \bibfield  {author} {\bibinfo {author} {\bibfnamefont {G.}~\bibnamefont
  {Henkelman}}, \bibinfo {author} {\bibfnamefont {B.~P.}\ \bibnamefont
  {Uberuaga}}, \ and\ \bibinfo {author} {\bibfnamefont {H.}~\bibnamefont
  {Jónsson}},\ }\href {\doibase 10.1063/1.1329672} {\bibfield  {journal}
  {\bibinfo  {journal} {The Journal of Chemical Physics}\ }\textbf {\bibinfo
  {volume} {113}},\ \bibinfo {pages} {9901} (\bibinfo {year}
  {2000})}\BibitemShut {NoStop}%
\bibitem [{\citenamefont {Togo}\ and\ \citenamefont {Tanaka}(2015)}]{phonopy}%
  \BibitemOpen
  \bibfield  {author} {\bibinfo {author} {\bibfnamefont {A.}~\bibnamefont
  {Togo}}\ and\ \bibinfo {author} {\bibfnamefont {I.}~\bibnamefont {Tanaka}},\
  }\href {\doibase https://doi.org/10.1016/j.scriptamat.2015.07.021} {\bibfield
   {journal} {\bibinfo  {journal} {Scripta Materialia}\ }\textbf {\bibinfo
  {volume} {108}},\ \bibinfo {pages} {1} (\bibinfo {year} {2015})}\BibitemShut
  {NoStop}%
\bibitem [{\citenamefont {Noguchi}\ \emph {et~al.}(2021)\citenamefont
  {Noguchi}, \citenamefont {Kobayashi}, \citenamefont {Jiang}, \citenamefont
  {Kuroda}, \citenamefont {Takahashi}, \citenamefont {Xu}, \citenamefont {Lee},
  \citenamefont {Hirayama}, \citenamefont {Ochi}, \citenamefont {Shirasawa},
  \citenamefont {Zhang}, \citenamefont {Lin}, \citenamefont {Bareille},
  \citenamefont {Sakuragi}, \citenamefont {Tanaka}, \citenamefont {Kunisada},
  \citenamefont {Kurokawa}, \citenamefont {Yaji}, \citenamefont {Harasawa},
  \citenamefont {Kandyba}, \citenamefont {Giampietri}, \citenamefont {Barinov},
  \citenamefont {Kim}, \citenamefont {Cacho}, \citenamefont {Hashimoto},
  \citenamefont {Lu}, \citenamefont {Shin}, \citenamefont {Arita},
  \citenamefont {Lai}, \citenamefont {Sasagawa},\ and\ \citenamefont
  {Kondo}}]{BiX}%
  \BibitemOpen
  \bibfield  {author} {\bibinfo {author} {\bibfnamefont {R.}~\bibnamefont
  {Noguchi}}, \bibinfo {author} {\bibfnamefont {M.}~\bibnamefont {Kobayashi}},
  \bibinfo {author} {\bibfnamefont {Z.}~\bibnamefont {Jiang}}, \bibinfo
  {author} {\bibfnamefont {K.}~\bibnamefont {Kuroda}}, \bibinfo {author}
  {\bibfnamefont {T.}~\bibnamefont {Takahashi}}, \bibinfo {author}
  {\bibfnamefont {Z.}~\bibnamefont {Xu}}, \bibinfo {author} {\bibfnamefont
  {D.}~\bibnamefont {Lee}}, \bibinfo {author} {\bibfnamefont {M.}~\bibnamefont
  {Hirayama}}, \bibinfo {author} {\bibfnamefont {M.}~\bibnamefont {Ochi}},
  \bibinfo {author} {\bibfnamefont {T.}~\bibnamefont {Shirasawa}}, \bibinfo
  {author} {\bibfnamefont {P.}~\bibnamefont {Zhang}}, \bibinfo {author}
  {\bibfnamefont {C.}~\bibnamefont {Lin}}, \bibinfo {author} {\bibfnamefont
  {C.}~\bibnamefont {Bareille}}, \bibinfo {author} {\bibfnamefont
  {S.}~\bibnamefont {Sakuragi}}, \bibinfo {author} {\bibfnamefont
  {H.}~\bibnamefont {Tanaka}}, \bibinfo {author} {\bibfnamefont
  {S.}~\bibnamefont {Kunisada}}, \bibinfo {author} {\bibfnamefont
  {K.}~\bibnamefont {Kurokawa}}, \bibinfo {author} {\bibfnamefont
  {K.}~\bibnamefont {Yaji}}, \bibinfo {author} {\bibfnamefont {A.}~\bibnamefont
  {Harasawa}}, \bibinfo {author} {\bibfnamefont {V.}~\bibnamefont {Kandyba}},
  \bibinfo {author} {\bibfnamefont {A.}~\bibnamefont {Giampietri}}, \bibinfo
  {author} {\bibfnamefont {A.}~\bibnamefont {Barinov}}, \bibinfo {author}
  {\bibfnamefont {T.~K.}\ \bibnamefont {Kim}}, \bibinfo {author} {\bibfnamefont
  {C.}~\bibnamefont {Cacho}}, \bibinfo {author} {\bibfnamefont
  {M.}~\bibnamefont {Hashimoto}}, \bibinfo {author} {\bibfnamefont
  {D.}~\bibnamefont {Lu}}, \bibinfo {author} {\bibfnamefont {S.}~\bibnamefont
  {Shin}}, \bibinfo {author} {\bibfnamefont {R.}~\bibnamefont {Arita}},
  \bibinfo {author} {\bibfnamefont {K.}~\bibnamefont {Lai}}, \bibinfo {author}
  {\bibfnamefont {T.}~\bibnamefont {Sasagawa}}, \ and\ \bibinfo {author}
  {\bibfnamefont {T.}~\bibnamefont {Kondo}},\ }\href {\doibase
  10.1038/s41563-020-00871-7} {\bibfield  {journal} {\bibinfo  {journal}
  {Nature Materials}\ }\textbf {\bibinfo {volume} {20}},\ \bibinfo {pages}
  {473} (\bibinfo {year} {2021})}\BibitemShut {NoStop}%
\bibitem [{\citenamefont {Zhou}\ \emph {et~al.}(2014)\citenamefont {Zhou},
  \citenamefont {Feng}, \citenamefont {Liu}, \citenamefont {Guan},\ and\
  \citenamefont {Yao}}]{BiBr-1layer.nl}%
  \BibitemOpen
  \bibfield  {author} {\bibinfo {author} {\bibfnamefont {J.-J.}\ \bibnamefont
  {Zhou}}, \bibinfo {author} {\bibfnamefont {W.}~\bibnamefont {Feng}}, \bibinfo
  {author} {\bibfnamefont {C.-C.}\ \bibnamefont {Liu}}, \bibinfo {author}
  {\bibfnamefont {S.}~\bibnamefont {Guan}}, \ and\ \bibinfo {author}
  {\bibfnamefont {Y.}~\bibnamefont {Yao}},\ }\href {\doibase 10.1021/nl501907g}
  {\bibfield  {journal} {\bibinfo  {journal} {Nano Letters}\ }\textbf {\bibinfo
  {volume} {14}},\ \bibinfo {pages} {4767} (\bibinfo {year} {2014})},\ \bibinfo
  {note} {pMID: 25058154},\ \Eprint
  {http://arxiv.org/abs/https://doi.org/10.1021/nl501907g}
  {https://doi.org/10.1021/nl501907g} \BibitemShut {NoStop}%
\end{thebibliography}
%

\end{document}